\documentclass[aps,amsmath,superscriptaddress,twocolumn,footinbib]{revtex4-2}
\usepackage{graphicx}
\usepackage{dcolumn} 
\usepackage{bm}
\usepackage[utf8]{inputenc}
\usepackage{amsmath}
\usepackage{amsthm}
\usepackage{hyperref}
\hypersetup{colorlinks,allcolors=blue}
\usepackage{physics}
\usepackage{mymacros}
\usepackage{siunitx}
\usepackage{natbib}
\usepackage{amsfonts}
\usepackage{tikz,pgf}
\usepackage{tabularx,stackengine}
\usepackage{soul}

\newcommand{\rarg}{(\bm r)}
\newcommand{\zvc}{\d_{\rm Z}}
\renewcommand{\CC}{\mathcal{C}}
\renewcommand{\WW}{\mathcal{W}}

\newcommand{\II}{\mathcal{I}}
\newcommand{\PP}{\mathcal{P}}
\renewcommand{\SS}{\mathcal{S}}
\newcommand{\Z}{\mathbb{Z}}
\newcommand{\pbar}{\bar{\partial}}
\newcommand{\Th}{\Theta}
\newcommand{\Si}{\Sigma}

\begin{document}
\title{Symmetries as the guiding principle for flattening bands of Dirac fermions}
\author{Yarden Sheffer}
\email{yarden.sheffer@gmail.com}
\affiliation{Department of Condensed Matter Physics, Weizmann Institute of Science Rehovot 7610001, Israel}
\author{Raquel Queiroz}
\email{raquel.queiroz@columbia.edu}
\affiliation{Department of Condensed Matter Physics, Weizmann Institute of Science Rehovot 7610001, Israel}
\affiliation{Department of Physics, Columbia University, New York, NY, USA}
\author{Ady Stern}
\email{adiel.stern@weizmann.ac.il}
\affiliation{Department of Condensed Matter Physics, Weizmann Institute of Science Rehovot 7610001, Israel}
\begin{abstract}
Since the discovery of magic-angle twisted bilayer graphene (TBG), flat bands in Dirac materials have become a prominent platform for realizing strong correlation effects in electronic systems. Here we show that the symmetry group protecting the Dirac cone in such materials determines whether a Dirac band may be flattened by the tuning of a small number of parameters. We devise a criterion that, given a symmetry group, allows for the calculation of the number of parameters required to make the Dirac velocity vanish. This criterion is employed to study band flattening in twisted bilayer graphene and in surface states of 3D topological insulators. Following this discussion, we identify the symmetries under which the vanishing of the Dirac velocity implies the emergence of perfectly-flat bands. Our analysis allows us to construct additional model Hamiltonians that display perfectly-flat bands at certain points in the space of parameters: the first is a toy model of two coupled 3D TI surfaces, and the second is a quasi-crystalline generalization of the chiral model of TBG.
\end{abstract}
\maketitle

\section{Introduction}
Since the discovery of superconductivity and correlated insulating states in magic-angle twisted bilayer graphene (TBG) \cite{Bistritzer2011a, PhysRevB.82.121407, Cao2018, Cao2018a} moir\'e materials have drawn tremendous attention as a tunable platform for creating novel electronic effects. The main feature of TBG is that by tuning the twist angle between the graphene layers one can tune the Dirac velocity at the Dirac cones to vanish to a remarkable degree of precision. The vanishing of the Dirac velocity is accompanied by a large density of states (DOS) at charge neutrality, thereby enhancing correlation effects. Following the example of magic-angle TBG, similar fine-tuned systems were shown to exhibit band flattening, with some examples being twisted trilayer graphene \cite{Khalaf2019, Carr2020, PhysRevLett.125.116404, park2021tunable}, twisted superconductors \cite{volkov2020magic, can2021high}, and moir\'e patterns on the surfaces of 3D topological insulators (TI) \cite{cano2021moire,  wang2021moire, dunbrack2021magic}. 

The emerging plethora of flat-band Hamiltonians in fine-tuned materials raises the question of how generic this phenomenon is. In other words, what characterizes the set of systems for which fine-tuning a small set of parameters leads to the formation of flat bands or almost flat bands? This question is interesting both from the theoretical and a practical point of view, as a criterion for band flatness should be a useful guide in searching for new materials where exotic correlated phenomena may be found. 

In this work, we focus on flat bands in systems harboring Dirac fermions, and more specifically on the conditions for flattening a band by making the Dirac velocity vanish. This scenario is relatively convenient to analyze theoretically, as it requires the knowledge of the Bloch Hamiltonian at only a single $k$ point. The choice to focus on the Dirac velocities can also be motivated by noting that generically an upper bound to the bandwidth may be estimated by a Debye-like approximation to the band dispersion. More rigorously,
in many cases of interest (see the examples discussed below) the quadratic order of the band dispersion near the $k$ point vanishes by symmetry. In these cases, the vanishing of the Dirac velocity guarantees that the DOS diverges at least as $(\delta E)^{-1/3}$ near the Dirac point. Notice that a quadratic band-touching cannot be obtained when the Dirac points are fixed by the symmetries of the system (for example by a rotation symmetry) as a result of the $\p$ Berry phase of the Dirac cone \cite{Xiao2010}. 

We begin in Sec. \ref{sec: zero velocity codimension} by defining an algebraic criterion: We show that the symmetry group $G$ acting on the Dirac cones determines the number of parameters (e.g., twist angle, pressure, etc...) that should generically be tuned to make the Dirac velocities vanish. This criterion is used to analyze different symmetry groups which can protect a Dirac cone, and to find the classes which allow for the tuning of a small number of parameters to obtain vanishing Dirac velocities. 

In Sec. \ref{sec: applications of zvc} we apply our criterion to the analysis of two systems of interest: The first one is band flattening of TBG, where we show that the existence of an approximate particle-hole symmetry is necessary for the vanishing of the Dirac velocity. The second system is surface-states of 3D TIs under a periodic potential. We show for such systems that the Dirac velocity at charge neutrality can be made to vanish entirely by varying a $C_2$ symmetric potential. The resulting system might enable a platform for realizing strongly-interacting phases on the surface of a TI. 

The vanishing of the Dirac velocity may be the first step towards a further increase of the DOS that culminates in a perfectly flat band \cite{Tarnopolsky2019, PhysRevB.103.165113, PhysRevB.103.155150}. The first example of a model that exhibits such a band was a toy model of TBG \cite{Tarnopolsky2019}. In Sec. \ref{sec: Exactly-flat bands} we extend the ideas raised by \cite{PhysRevB.103.155150} and our discussion of the vanishing Dirac velocity to discuss the symmetry requirements that are needed to obtain exactly-flat bands in general settings. We show that such flat-band Hamiltonians naturally arise in Dirac Hamiltonians with an external $SU(2)$ gauge field by tuning a small number of parameters. For Hamiltonians in class CI of the Altland-Zirnbauer classification \cite{Altland1997} we show that the flat-bands condition is equivalent to the vanishing of the Dirac velocity. For Hamiltonians with more general symmetries, we show that the flat bands can be found by considering the vanishing of the velocity in a modified version of the original Hamiltonian, which is in class CI. 

We employ our discussion of exactly flat bands in Sec. \ref{sec: applications of flat band models} where we discuss two new model Hamiltonians which realize such exactly-flat bands, along with an in-depth analysis of a recently-proposed model. The first example is a continuum model with a $C_4$ symmetry, which can be thought of as a toy model of two TI surfaces with spin-flipping tunneling and an in-plane position-dependent magnetic field. The second example is a model of a quadratic band-touching Hamiltonian first proposed by \cite{PhysRevResearch.4.043151}, on which our analysis can be used to prove analytically the existence of exactly-flat bands. The last model is a quasi-crystalline generalization of the chiral TBG Hamiltonian. While the latter model does not have well-defined bands, we show it to host ``magic angles" with an extensive degeneracy at charge neutrality.

Sec. \ref{sec: discussion} concludes with a discussion of possible future directions. The appendices contain a more rigorous definition of our algebraic criterion, reviews of known results published elsewhere, technical proofs, and a discussion of edge cases that are not treated in the main text.
\section{Conditions for the vanishing of the Dirac velocity} 
\label{sec: zero velocity codimension}
\subsection{Zero-velocity co-dimension}
\begin{figure}
	\centering
	\includegraphics[scale=1]{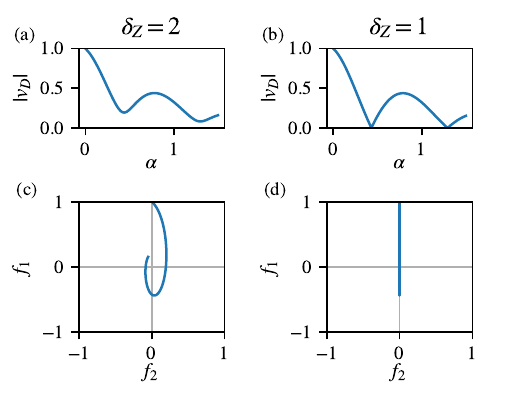}
	\caption{\label{fig:vx_trajectory} The trajectory of $\r(v_x)$ for a varied parameter $\a$. (a),(c) depict the trajectory of $\qty|v_x|$ and $\r(v_x)$ ($f_{1,2}$ are defined in \eqref{eq:v_decomposition}). Since the trajectory in $f_{1,2}$ does not cross zero the Dirac velocity does not vanish. (b),(d) are similar but with $\zvc=1$. The additional constraint on $f_{1,2}$ allows $\qty|v_x|$ to vanish on certain values of $\a$.  }
\end{figure}
Consider a two-dimensional Bloch Hamiltonian $H(\bm k)$ whose band structure has Dirac points for certain values of $\bm k$. The velocity operators at the Dirac points are defined by 
\begin{equation}
\label{eq: velocity operator}
	v_i = \frac{\partial H(\bm k)}{ \partial k_i}.
\end{equation}
The Dirac velocities (that is, the dispersion of the Dirac cone close to the Dirac point) are calculated using first-order perturbation theory of $\bm v$ acting on the degenerate wavefunctions at the Dirac cone. They are the eigenvalues of the matrices
\begin{equation}
	\r\qty(v_i)_{mn} = \mel{\y_m}{v_i}{\y_n},
\end{equation}
where $\y_m$ are the degenerate Bloch wavefunctions at the Dirac cone. We will use the notation $\rho({\hat O})$ to denote the projection of the operator ${\hat O}$ onto the subspace spanned by $\{\psi_m\}$. Most commonly $m=1,2$, but we shall also consider the cases of $n_D$ degenerate Dirac cones, for which $m=1,...,2n_D$. We note that when the degenerate Dirac cones are protected by a local unitary symmetry (such as $SU(2)$ spin rotation or a translation symmetry), we can consider each eigenspace of the symmetry separately as a single Dirac cone. 

Our main interest is the condition for the Dirac velocities to vanish at some points in the space of parameters. We assume that $H$ is controlled by a set of $d$ parameters $\a_1,...,\a_d$. We define the \textit{zero velocity codimension} $\zvc$ to be the codimension of the manifold in the space of $\alpha_i$ for which $\rho\left(v_x\right)=\r\qty(v_y)=0$. Roughly speaking, if the Hamiltonian $H$ can have a vanishing Dirac velocity, $\zvc$ is the number of parameters that should be tuned to make the velocity vanish. Our goal in this section is to show how $\zvc$ can be calculated from the symmetries that preserve the Dirac cone. 

Let $G$ be the group of symmetries that preserve the Dirac cone. For such symmetries $\rho(G)$ is a representation of $G$, for $g\in G$ being a unitary operator. In the case where $g$ is an antiunitary operator we obtain an antiunitary representation by multiplying the matrix $\r(g)_{mn}$ by the complex-conjugation operator $K$. Since the elements $g$ relate states $\ket{\y_i}$ only to one another, they satisfy, 
\begin{equation}
	\label{eq:rep_comm}
	\r(g)\r\qty(v_i)\r(g)^{-1} = \r(gv_i g^{-1})
\end{equation}
for all $g\in G$. Note that lattice symmetries can relate $v_x$ and $v_y$. The tuples $\qty(\r(v_x),\r(v_y))$ are therefore elements of the linear subspace $V$ of tuples of $2n_D$-dimensional Hermitian matrices that satisfy \eqref{eq:rep_comm}. We have 
\begin{equation}
	\label{eq:v_decomposition}
	\begin{pmatrix}
	\rho(v_x) \\ \rho(v_y)
	\end{pmatrix} = \sum_{l=1}^{\zvc} f_l(\a_1,...,\a_d) \begin{pmatrix}
	M_{l,x} \\ M_{l,y}
	\end{pmatrix}
\end{equation}
where $f_l(\a_1,...,\a_d)$ are real-valued and $(M_{l,x},M_{l,y})$ give a basis for $V$. Consequently, $\zvc$ is the dimension of $V$. Eq. \eqref{eq:v_decomposition} is then the statement of how the symmetries of the Dirac cones define $\zvc$. In the absence of symmetries protecting the Dirac cones, \eqref{eq:v_decomposition} means trivially that the matrices $\r(v_i)$ can be expanded in a basis of Hermitian matrices, and $\zvc$ is the dimension of all possible tuples of such matrices, given by $8n_D^2$ (for example, with $n_D=1$, $M_{i}$ could be any of $\s_{0,x,y,z}$, giving a total of 8 tuples).

In the case of  $\zvc=1$, Eq. \eqref{eq:v_decomposition} reduces to
\begin{equation}
	\label{eq:v_decomposition_d=1}
	\begin{pmatrix}
	\r\qty(v_x) \\\r\qty(v_y)
	\end{pmatrix}= f(\alpha_1,...,\alpha_n) 
	\begin{pmatrix}
	M_x \\ M_y
	\end{pmatrix}. 
\end{equation}
$\rho\qty(v_i)$ are then fixed up to a real parameter and we obtain a vanishing Dirac velocity whenever $f$ vanishes (see Fig. \ref{fig:vx_trajectory}). In that case, we can make the Dirac velocity vanish exactly by tuning a single parameter. In Fig. \ref{fig:vx_trajectory} we show the trajectory for $\r(v_x)$ for $\zvc=2$ and $\zvc=1$ as we vary a single parameter. Noticeably, the Dirac velocity can vanish exactly as $\a$ is varied only when $\zvc=1$. In general, given that we have $d$ parameters and $\zvc$ equations, the dimension of the zero-velocity solutions is $d-\zvc$. Notice that, since $f$ in \eqref{eq:v_decomposition_d=1} is a continuous function, a change of sign in the Dirac velocity implies the existence of a point where it vanishes. Also, this sign is in agreement with the sign of $v_D$ obtained in perturbation-theory, e.g. in \cite{Tarnopolsky2019}.

A slightly more rigorous definition of $\zvc$ is given in Appendix \ref{app:rigorous_zvc}. We show there that if the Dirac velocity is made to vanish at some point $\vec{\a}_0$ in the parameters space, there exists a manifold of dimension $d-\zvc$ around $\vec{\a}_0$ in parameter space where the Dirac velocity remains zero. The proof relies on the assumption that the gap between the Dirac point and the rest of the bands does not close. Such closing of the gap results in the Dirac cone wavefunctions not being continuous and can create a boundary to the zero-velocity manifold. We treat an example of such gap closing in Appendix \ref{app: C_4_breaking}.

\subsection{Calculation of $\zvc$ for different symmetry groups}
We now follow the principles outlined above to calculate $\zvc$ for different symmetry groups $G$ which preserve the Dirac point. The symmetry groups we choose to focus on may contain two antiunitary symmetries $\Theta,\Pi$ that anticommute ($\Theta$) and commute ($\Pi$), with the operators $v_i$, as well as their unitary product $\Sigma$, 
\begin{equation}
\label{eq: TPS_commutators}
    \qty{ v_i,\Theta} =\qty[v_i,\Pi]=\qty{ v_i,\Sigma}  =0.
\end{equation} 
In cases where the system has local time-reversal $T$ and particle-hole $P$ symmetries that map the Dirac cone onto itself, they may serve as $\Theta$ and $\Pi$ respectively. Such is the case for a Dirac cone on the surface of a three-dimensional TI. When $T, P$ map between different Dirac cones, such as in the case of TBG, we can combine them with other unitary symmetries to form $\Theta,\Pi$. We will identify these combinations when we discuss examples of the latter case. In general, we do not demand that the symmetries are local. Furthermore, while we assume that the symmetries either commute or anticommute with the Hamiltonian, we define them only by the commutation relations \eqref{eq: TPS_commutators} and not by their commutation/anticommutation relations with the Hamiltonian.

When the symmetry group $G$ exists, the symmetries constrain the possible representations of the two components of the velocity operator $(v_x,v_y)$. However, as long as the relations \eqref{eq: TPS_commutators} do not distinguish between $v_x$ and $v_y$, they are not enough to constrain $\zvc$ to $1$, since for any $M_l=(M_x,M_y)$ that satisfies them $\Tilde{M}_l=(M_y,-M_x)$ will do as well, leading to $\zvc\ge2$. To find cases for which $\zvc=1$ we need an additional symmetry that acts differently on $v_x$ and $v_y$. We, therefore, consider also a (unitary) reflection symmetry $R$ which takes $x\to-x$ and thus satisfies
\begin{equation}
\label{eq:R_commutators}
    \qty{R,v_x}=\qty[R,v_y]=0.
\end{equation}

Our main result in this section is a calculation of $\zvc$ for all symmetry groups constructed from $\Theta,\Pi,\Sigma,R$. Since $\Theta,\Pi$ are antiunitary, different symmetry groups are given by different choices of $\xi_{T,P}=\pm1$ defined by
\begin{equation}
\label{eq:xi_definition}
    \begin{aligned}
    \Th^2 &=\xi_\Th \\
    \Pi^2 &= \xi_\Pi.
    \end{aligned}
\end{equation}

Besides the sign choice of $\xi_{\Pi,\Si}$, different symmetry groups are distinguished by allowing $R$ to either commute or anticommute with $\Theta,\Pi,\Sigma$. Namely, for $R$ fixed by $R^2=+1$ we can choose
\begin{equation}
\label{eq:zeta_definition}
    \begin{aligned}
        R\Th &= \zeta_\Th \Th R \\
        R\Pi &= \zeta_\Pi \Pi R \\
        R\Si &= \zeta_\Si \Si R.
    \end{aligned}
\end{equation}
Again $\zeta_{\Th/\Pi/\Si}=\pm1$ and, assuming that all are present, $\zeta_\Si=\zeta_\Th\zeta_\Pi$. In the case where we have both $\Th,\Pi$ we use the notation $R_{\zeta_\Th\zeta_\Pi}$ to denote the commutation/anticommutation relations, while a single subscript will be used in the classes where we have only one of $\Pi,\Th,\Si$. The resulting family of symmetry groups is similar to that considered in the Altland-Zirnbauer classification with reflection symmetry \cite{Altland1997, moore2014quantum, Morimoto2013, chiu2013classification, chiu2016classification}. 

We calculate $\zvc$ for $n_D=1,2$.  Table \ref{tab:maintext_zvc} presents the results for $n_D=1$. Table \ref{tab:single DC} presents the same case with more details and table \ref{tab:Double DC} presents the results for $n_D=2$ (Tables \ref{tab:single DC} and \ref{tab:Double DC} are presented in the Appendices). The tables are constructed as follows: Assuming that the Dirac cone is $n_D$-times degenerate, for each symmetry group we construct a representation $\rho$. We then look for the possible representations of $v_{x}, v_y$ which satisfy the commutation relations with the symmetry operators obtained from \eqref{eq: TPS_commutators},\eqref{eq:R_commutators}. Finally, we assume the presence of a crystalline symmetry relating $v_x,v_y$ such that $\zvc$ is determined only by the dimension of possible $\r(v_x)$. Examples of such symmetry are $C_3$ and $C_4$ symmetries, where $C_n$ is a rotation of the system by $2\p/n$. 

A similar analysis can be straightforwardly extended to $n_D$-fold degenerate Dirac cones for higher $n_D$, include additional symmetries, or extend to three-dimensional Weyl and Dirac nodes \cite{RevModPhys.90.015001}.
\begin{table}
    \centering
    \begin{tabular}{|cccccc|}
    \hline 
    $\ \Theta\ $ & $\ \Pi\ $ & $\ \Sigma\ $ & $\ R\ $  & $\ \zvc\ $ & Example\tabularnewline
    \hline 
    \hline 
    0 & 0 & 0 & $R$ & 2 & \tabularnewline
     &  &  & 0 & 3 & \tabularnewline
    0 & 0 & 1 & $R_{-}$ & 1 & \scriptsize{Nodal SC w/ crystalline symmetries }\tabularnewline
     &  &  & 0 & 2 & \scriptsize{Nodal SC}\tabularnewline
    0 & + & 0 & $R_{+}$ & 1 & \scriptsize{TBG (approx.)}\tabularnewline
     &  &  & 0 & 2 & \scriptsize{TBG (exact), near-commensuration TBG}\tabularnewline
    - & + & 1 & $R_{-+}$ & 1 &  \scriptsize{$\mathbb{Z}_{2}$ TI surface states w/ crystalline symmetries}\tabularnewline
     &  &  & 0 & 2 & \tabularnewline
    - & 0 & 0 & $R_{-}$ & 2 & \tabularnewline
     &  &  & 0 & 3 & \scriptsize{$\mathbb{Z}_{2}$ TI surface states (general)}\tabularnewline
    \hline 
    \end{tabular}
    \caption{$\zvc$ for the symmetry groups (with $\Theta,\Pi,\Sigma$ and/or $R$ symmetries) which preserve a single Dirac cone. Zero indicates the absence of a symmetry, while $\pm$ denote the square of the symmetry. 
    The signs at the subscripts of $R$ indicate whether $R$ commutes/anticommutes with the other symmetries. The calculation of $\zvc$ here assumes the presence of a $C_3$ or $C_4$ symmetry. See Table \ref{tab:single DC} for an elaborated calculation of $\zvc$. See  \cite{volkov2020magic}  for a discussion of the example in row 3, Sec. \ref{sec: applications of zvc}A for a discussion of rows 5 and 6, and Sec. \ref{sec: applications of zvc}B for discussion of row 7.}
    \label{tab:maintext_zvc}
\end{table}
\section{Applications}
\label{sec: applications of zvc}
\subsection{Magic angles in TBG}
As a first application of our analysis, let us consider the band flattening in TBG. Since the two Dirac points are fixed in different $ k$ points, we calculate $\zvc$ for a single Dirac cone. Time reversal $T$ maps between the two valleys of the electronic band. However, when multiplied by $C_2$, we obtain an antiunitary symmetry that preserves the Dirac point and commutes with the velocity operators, such that it may serve as $\Pi$. Furthermore, the  $C_3$ symmetry preserves the Dirac point as well (see a review of the TBG Hamiltonian and symmetries in Appendix \ref{app: MATBG}). We fix the representation of these symmetries on the two Dirac point wavefunctions to be
\begin{equation}
	\begin{aligned}\r\left(C_{3}\right) & =e^{i\frac{2\p}{3}\s_{z}},\\
		\r\left(C_{2}T\right) & =\s_{x}K
	\end{aligned}
\end{equation}
which is dictated by the requirements that $\r\left(C_{2}T\right)$ should be anti-unitary, should square to $+1$, and should satisfy  $\r(C_3)\r(C_2T)\r(C_3)^{-1}=\r(C_2T)$. For the Dirac cone to be $C_3$ symmetric we must have $\{v_x,v_y\}=0$ such that \eqref{eq:v_decomposition} becomes
\begin{equation}
	\begin{aligned}
		\r\left(v_{x}\right) & =f_{1}\left(\a\right)\s_{x}+f_{2}\left(\a\right)\s_{y}, \\
		\r\left(v_{y}\right) & =f_{1}\left(\a\right)\s_{y}-f_{2}\left(\a\right)\s_{x}.
	\end{aligned}
\end{equation}
Evidently,  these  two symmetries are not sufficient to ensure that the Dirac velocity may be made to vanish with a variation of a single parameter $\alpha$. Luckily, TBG at small twist angle $\q$ has an additional \textit{approximate} unitary particle-hole symmetry (broken by a term of order $O(\q)$) given by \cite{Song2019}
\begin{equation}
\begin{aligned}
    \mathcal{C}:& &\h_y \s_x K.
    \end{aligned}
\end{equation}
where $\h_i$ are the Pauli matrices acting on the layer indices. This symmetry can be combined with the exact symmetry $C_{2,x}$ to form an additional symmetry that preserves the Dirac cone
(see Appendix \ref{app: MATBG} for a review of the symmetries of TBG). Thus, under the approximation of a small twist angle the operator $\mathcal{C}C_{2,x}$ maps $x\rightarrow -x$ and anticommutes with the Hamiltonian at low energies. Consequently, it commutes with $v_x$ and anticommutes with $v_y$. Choosing its representation to be $\sigma_x$, it fixes $f_2=0$. Consequently, the magnitude of the Dirac velocity is given by $\qty|f_1(\a)|$, which may be made zero when $f_1$ changes sign. 

\begin{figure}
    \centering
    \includegraphics[scale=.72]{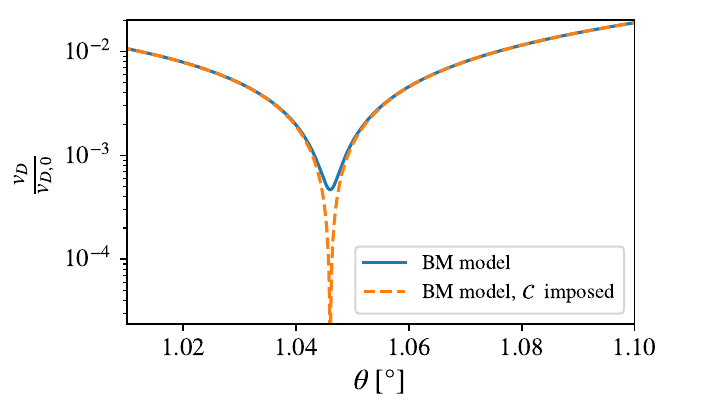}
    \caption{The Dirac velocity $v_D$ of the BM model \eqref{eq:BM_H}, with and without the additional $\CC$ symmetry, as the twist angle $\theta$ is varied near the magic angle. The $\CC$ symmetry is imposed by setting $\theta=0$ in the $h(\theta)$ terms.}
    \label{fig:tbg_velocity}
\end{figure}

This calculation leaves us with an important lesson: in TBG, both the exact and the approximate symmetries are necessary for the Dirac velocity to vanish at the magic angle. Indeed, by diagonalizing the Bistritzer-MacDonald (BM) Hamiltonian \cite{Bistritzer2011a}, we find that when one does not impose the approximate symmetry to be exact the Dirac velocity does not reach zero when $\q$ is varied. It instead has a minimum value of $\approx4\times10^{-4}\times v_0$ where $v_0$ is the Dirac velocity at zero coupling between the layers. The value of $v_D$ as the twist angle is varied, with and without the approximate $\mathcal{C}$ symmetry, is depicted in Fig. \ref{fig:tbg_velocity}.

In magic-angle TBG the $\mathcal{C}$ symmetry is weakly broken by a symmetry-breaking term proportional to the small twist angle $\theta$.  When the twist angle is not small, for example when the layers are slightly twisted away from an angle of commensuration \cite{scheer2022magic}, the symmetry-breaking term has a non-trivial dependence on the twist angle. In particular, it does not vanish at commensuration angles. In a recent work \cite{scheer2022magic}, the authors describe the case of twisted graphene bilayers when the layers are twisted slightly away from a commensurate twist angle. The Hamiltonian obtained is similar to the BM Hamiltonian, but the $\mathcal{C}$-symmetry is broken by a parameter that is independent of the deviation from the commensurate angle. As observed there, the Dirac velocity does not reach zero. Our analysis shows this to be a result of the breaking of $\mathcal{C}$ symmetry.
\subsection{$\zvc$ in 3D TI surface-states}
\begin{figure}
    \centering
    \includegraphics[scale=.8]{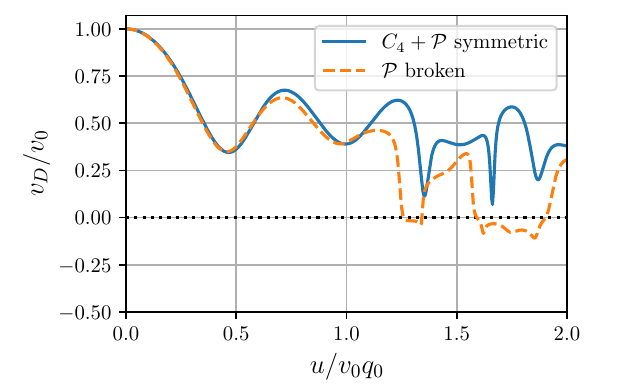}
    \caption{The Dirac velocity of the Hamiltonian \eqref{eq: TI_hamiltonian} with a $C_4$ symmetric potential in the $\PP$-symmetric and $\PP$ broken case. The $\PP$-symmetric case is defined by the potential \eqref{eq:TI_potential} with $u=u_x=u_y$ while the $\PP$ broken plot is given for the potential \eqref{eq:TI_p_broken_potential} with $u_1=u,u_2=\frac{u}{4}$. In the latter case, the velocity is defined as the velocity of the Dirac cone connected adiabatically to the Dirac cone at zero energy as $u$ is increased. One can see that $v_x$ can reach 0 when the $\PP$ antisymmetry is broken, but not when it is present. }
    \label{fig:TI_vel}
\end{figure}
Refs. \cite{cano2021moire, wang2021moire} suggested that the velocity characterizing the Dirac cone of the surface of a 3D TI may be suppressed by the application of a periodic potential on the surface. 
Here we use our analysis of $\zvc$ to show that there exist ``magic parameters" leading to an exact vanishing of the Dirac velocity. We present two types of periodic potentials that lead to a vanishing Dirac velocity: the first possesses a $C_4$ symmetry and requires tuning a single parameter. The second has only a $C_2$ symmetry, so that each of $v_x,v_y$ can be made to vanish by tuning a single parameter. The velocity vector can then be made to vanish entirely by tuning two parameters. 

The Dirac cone on the surface of a 3D TI is protected by time-reversal symmetry $T=\s_yK$. A periodic potential is consistent with this symmetry, leading to the Hamiltonian of the form
\begin{equation}
\label{eq: TI_hamiltonian}
    \HH=v_0\bm \s \cdot \bm p +u\rarg
\end{equation}
where $v_0$ is the Dirac velocity at zero external potential and $u\rarg$ is a periodic potential term. Note that $T$ allows only for a potential term proportional to the identity in \eqref{eq: TI_hamiltonian} and prohibits the opening of a gap. 

In the case where, in addition to $T$, there exist additional $C_n$ ($n=4,6$) and $M_x=\s_y (x\to-x)$ crystalline symmetries \footnote{Note that spinful reflection symmetry is given by $M_x=\s_x(x\to-x)$. Here we choose a gauge in which the Dirac cone Hamiltonian is of the form $\bm k\cdot\bm \s$ which gives the present definition of $M_x$.}, we find from Table \ref{tab:maintext_zvc} that $\zvc=1$ (since the Dirac point maps to itself under time reversal we have $\Theta=T$,$\Sigma=C_2,R=M_x$). 

When a $C_2$ symmetry is present in the system, $\r(v_i)$ have no diagonal terms in the basis defined by $\ket{\y},T\ket{\y}$, provided that $\ket{\y}$ is a Dirac-cone wavefunction that is a $C_2$ eigenfunction. The Dirac velocity can then be calculated by 
\begin{align}
        v_x&=\mel{T\y}{v_0\s_x}{\y}=v_0\mel{K\y}{\s_z}{\y} \\
        v_y&=\mel{T\y}{v_0\s_y}{\y}=v_0\braket{K\y}{\y}.
\end{align}
To make the discussion concrete, we start by taking $u\rarg$ of the form
\begin{equation}
    \label{eq:TI_potential}
    u\rarg=2u_x \cos{q_0 x}+2u_y \cos{q_0 y}.
\end{equation}
Generally, the Hamiltonian \eqref{eq: TI_hamiltonian} with \eqref{eq:TI_potential} is symmetric under $C_2=\s_z(\bm r \to - \bm r)$ and $M_x=\s_y (x\to - x)$. When $u_x=u_y$ it is also symmetric under $C_4=e^{i\p/4 \s_z}(\bm r \to \mathcal{R}_4\bm r)$. Furthermore, the Hamiltonian anticommutes with the operator
\begin{equation}
\label{eq:TI_p_broken_potential}
    \mathcal{P}=\s_z (x\to x+\p/q_0,y\to y+\p/q_0)
\end{equation}
We now analyze the system both in the case where there is an additional $C_4$ symmetry and where this symmetry is broken. 

\subsubsection{$C_4$ symmetric case}
Our numerical studies of the $C_4$-symmetric case indicate that the anticommutation of $H$ and $\PP$ prevents the vanishing of the Dirac velocity. Indeed, by diagonalizing the Hamiltonian we do not find any magic values (see Fig. \ref{fig:TI_vel}) as $u=u_x=u_y$ is varied. Besides the calculation presented in the figure, we checked that the Dirac velocity does not vanish up to $u=10$. We also find numerically that the velocity does not vanish even when one considers additional $C_4$ and $\PP$ preserving terms in $u\rarg$ with higher wave vectors. We therefore conjecture, but cannot prove generally, that a Hamiltonian of the form \eqref{eq: TI_hamiltonian} with $C_4,M_x$, and $\PP$ symmetries cannot yield a vanishing Dirac velocity for the Dirac cone at charge neutrality. 

One can break $\PP$ by introducing higher wave vectors in the potential $u$. As an example we take
\begin{equation}
\begin{aligned}
    u\rarg&=2u_1\qty( \cos{q_0 x}+ \cos{q_0 y})\\
    &+2u_2(\cos(q_0x+q_0y)+\cos(q_0x-q_0y)).
\end{aligned}
\end{equation}
When $\PP$ is broken, the Dirac cone, which for $u\rarg=0$ is at $E=0$ ceases to be fixed in energy. We can nevertheless calculate the velocity in the Dirac cone connected adiabatically to the one at $E=0$ as the amplitude of $u\rarg$ increases. As an example, in Figure \ref{fig:TI_vel} we plot the Dirac velocity for $u_1=u,u_2=u/4$ and find points along the line in which the velocity vanishes. Note that while the first-order dispersion around the Dirac cone vanishes, we still have quadratic terms in the dispersion, leading to a finite (but increased) density of states at the Dirac cone.

In this example, breaking the $\PP$-symmetry opened a way for the Dirac velocity to vanish, despite the insensitivity of $\zvc$ to this breaking. This may indicate that symmetries might also have an impeding role in the tuning of systems parameters to make the Dirac velocity vanish. Indeed, our analysis of $\zvc$ gives necessary conditions, but not sufficient conditions,  for making the Dirac velocity vanish by tuning a given number of parameters.

\begin{figure}
    \centering
    \includegraphics[scale=.85]{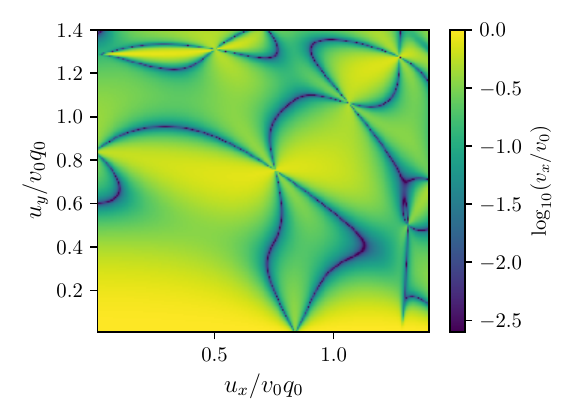}
    \caption{The $x$ component $v_x$ of the velocity for the Dirac cone at charge neutrality for the Hamiltonian \eqref{eq: TI_hamiltonian},\eqref{eq:TI_potential} (in logarithmic scale). One sees that $v_x$ vanishes on lines in the $u_x,u_y$ space, giving rise to a low energy Hamiltonian of the form \eqref{eq:quasi_1d}.}
    \label{fig:c4_broken_vx}
\end{figure}

\subsubsection{$C_4$-broken case: vanishing of the velocity in a single direction}

When the periodic potential breaks $C_4$ symmetry, the difficulties of making the Dirac velocity vanish are alleviated. In this case, each velocity component $v_{x,y}$ vanishes on a codimension-one manifold, which results in $\zvc=2$. Indeed we find lines of ``magic parameters" $u_x,u_y$ which give one vanishing component of the velocity at charge neutrality (see Fig. \ref{fig:c4_broken_vx}). By tuning both parameters, we find points at which both components of the velocity vanish. 

A simple, analytically-solvable example of this scenario can be found in the case where $u_y=0$ and the potential is one-dimensional. In that case, we can find zero-energy states of the form
\begin{align}
    \y_\pm(x)&=\sqrt{\frac{q_0}{8\p}}\begin{pmatrix}
    e^{ iU(x)}\pm e^{ -iU(x)}\\ e^{ iU(x)}\mp e^{ -iU(x)}
    \end{pmatrix}, \\ 
    U(x)&=\frac{2u_x}{q_0v_0}\sin(q_0 x),
\end{align}
The Dirac velocities can be obtained from the representations of the velocity operators in this basis. We obtain
\begin{align}
    \r(v_x)&=v_0 \s_x ,\\
    \label{eq:1d_potential_v_y}
        \r(v_y)&=v_0 \s_y J_0\qty(\frac{4 u_x }{ q_0 v_0})
\end{align} 
where $J_0$ is the Bessel function of the first kind. We notice the somewhat surprising result that it is $v_x$, and not $v_y$, which is independent of the potential \cite{katsnelson2006chiral}, even though the potential varies along the $x$-direction. The matrix $\r(v_y)$ is proportional to $\s_y$ as a result of the inversion symmetry of $u(x)$ and vanishes on the zeros of $J_0$. We then find the ``magic parameters" at $u_x/q_0v_0=.60,1.38,2.16,...$.

A velocity that vanishes only in the $y$ direction  gives rise to a low-energy Hamiltonian of the form
\begin{equation}
    \label{eq:quasi_1d}
    H=\tilde{v}_x \s_x k_x + \qty(\tilde{d}_x k_x^2+\tilde{d}_y k_y^2) k_y \s_y.
\end{equation}
for some parameters $\tilde{v}_x,\tilde{d}_x,\tilde{d}_y$. The DOS resulting from \eqref{eq:quasi_1d} vanishes as $g(E)\propto E^{1/3}$ at low energies. An interesting question, which will not be elaborated on here, is the behavior of the Hamiltonian \eqref{eq:quasi_1d} when interactions are also considered. Renormalization-group analysis \cite{wen1990metallic,  giamarchi2003quantum} suggests that in the presence of strong enough interactions, the dynamics of the system become quasi-one-dimensional, forming a Luttinger liquid phase in the $x$ direction \cite{ wang2021one}, possibly with spontaneous breaking of translation symmetry in the $y$-direction. 

\subsubsection{$C_4$-broken case: vanishing Dirac velocity in both directions,  with $\zvc=2$}
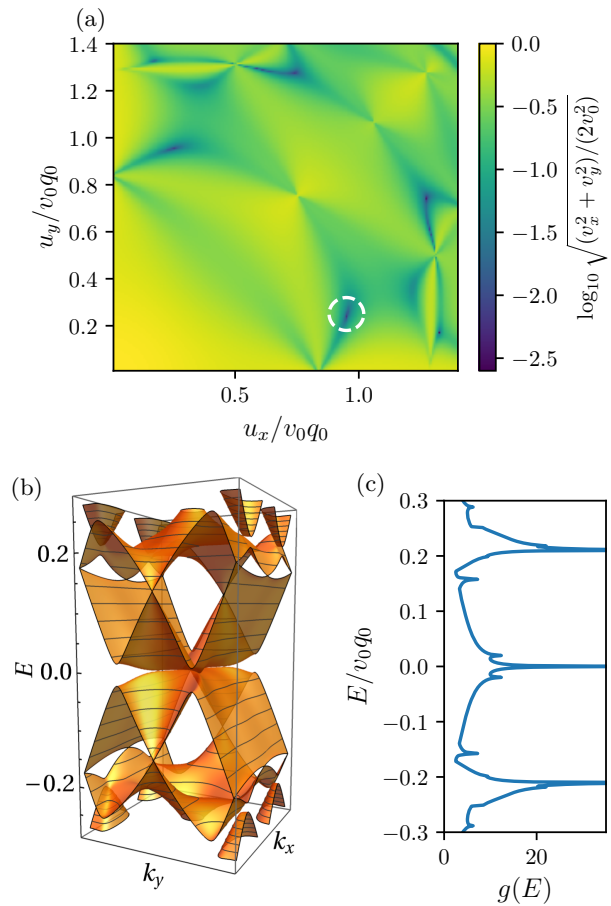
\begin{figure}
    \centering
    \begin{tikzpicture}
        \node at (0,0)
        {\includegraphics[scale=.85]{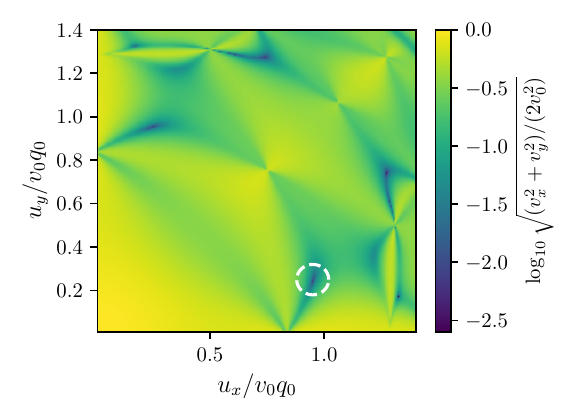}};
        \node at (-3,2.8) {(a)};
    \end{tikzpicture}
    \hfill
    \raisebox{.2cm}{
    \begin{tikzpicture}
        \node at (0,0)
        {\includegraphics[scale=.65]{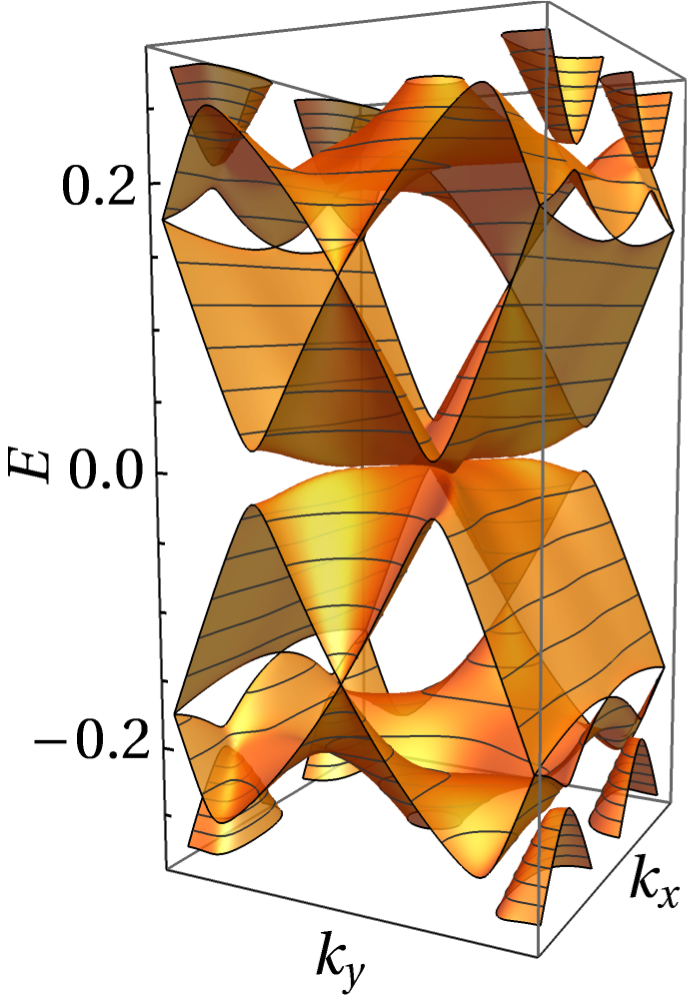}};
        \node at (-1.7,2.6) {(b)};
    \end{tikzpicture}}
    \begin{tikzpicture}
        \node at (0,0)
        {\includegraphics[scale=.85]{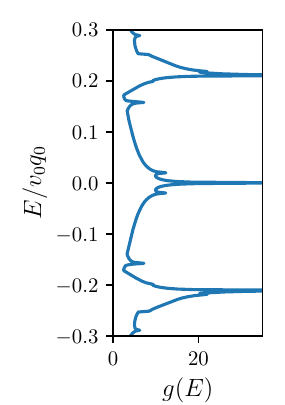}};
        \node at (-1.4,2.7) {(c)};
    \end{tikzpicture}
    
    \caption{(a) The ``absolute" Dirac velocity $\sqrt{v_x^2+v_y^2}$ of the Dirac cone at charge neutrality for the Hamiltonian \eqref{eq: TI_hamiltonian},\eqref{eq:TI_potential} as a function of $u_x,u_y$, in logarithmic scale. The dark spots are codimension-two manifolds on which the Dirac velocity vanishes entirely. (b,c) Band structure and DOS of the Hamiltonian \eqref{eq: TI_hamiltonian},\eqref{eq:TI_potential} at the ``magic angle" obtained with $u_x=0.95,u_y=0.25$ (the point is marked by a circle in (a)). We find a divergent DOS at charge neutrality with additional Van Hove singularities at $E=\pm0.21$. }
    \label{fig:TI_vel_2d}
\end{figure}

Since each velocity component vanishes on lines in $(u_x,u_y)$ space, we expect the entire velocity to vanish on points in that space. Fig. \ref{fig:TI_vel_2d}a shows that this is indeed the case. In Fig. \ref{fig:TI_vel_2d}b,c, we plot the band structure and DOS at one of these points. One can observe a peak in the DOS at charge neutrality, with two additional peaks corresponding to Van Hove singularities (vHs) at non-zero energies. Note that the functional dependency of the two peaks is different: while the DOS at the vHs diverges as $-\log(|\d E|)$ (with $\delta E$ being the deviation from the singularity), the DOS diverges as $\qty|\delta E|^{-1/3}$ around charge neutrality. Thus, the points in parameter space where the Dirac velocity vanishes might provide good candidates for strongly-interacting states at charge neutrality. Notice that the divergence is functionally similar to the higher vHs discussed in \cite{wang2021moire}, but the low-energy Hamiltonian around the critical point is different \cite{yuan2019magic, yuan2020classification}. 

The authors of \cite{cano2021moire} propose two methods for realizing a Hamiltonian of the form \eqref{eq:TI_potential} on the surface of a TI, either by creating a moir\'e pattern on the surface or by posing a dielectric pattern on it \cite{forsythe2018band}. Here we note that the degree of tunability required to achieve the ``magic coupling values" obtained in the $C_2$-symmetric model can be achieved either by a $C_2$ symmetric dielectric pattern or by a potential generated by acoustic waves on the surface \cite{PhysRevLett.65.112, simon1996coupling}. Notice that the dimensionless parameter controlling the coupling strength is $u/q_0v_0$. Ideally, one would keep both $u,q_0$ high to mitigate disorder effects and to have a large range of momenta affected by the modulation. 

Also, we note that in \cite{dunbrack2021magic} the authors show that for two 3D TI surface states a spin-flipping tunneling term allows for the velocity to vanish. When only spin-independent tunneling between two such surfaces is allowed, the surface Hamiltonians can be written in four-component spinors as
\begin{equation}
    \HH_{\text{twisted-TI}}=v_0\bm k\cdot\bm \s +\eta_x u\rarg
\end{equation}
when $\eta_i$ are the ``surface" indices. The Hamiltonian $\HH_{\text{twisted-TI}}$, therefore, splits into layer symmetric and antisymmetric sections, each described by the Hamiltonian \eqref{eq:TI_potential}. Our findings then provide two additional mechanisms for obtaining a Dirac cone with vanishing velocity in such systems.

\section{Exactly-flat bands}
\label{sec: Exactly-flat bands}
An intriguing aspect of the BM Hamiltonian is the presence of a limit \cite{Tarnopolsky2019} in which the Hamiltonian has an additional chiral symmetry and in which the bands  at charge neutrality become perfectly flat at the magic angle. These bands may then be chosen to have non-zero Chern numbers and a well-defined sublattice polarization \cite{Bultinck2020}. The Hamiltonian at that limit, referred to as cTBG, allows for exactly flat bands to be reached by tuning a single parameter. In this section, we provide symmetry requirements under which the vanishing of the Dirac velocity implies that the bands are exactly flat. Our analysis provides conditions for small-codimension exactly-flat bands. 

We begin by considering a generalized form of the cTBG Hamiltonian, described by a Dirac electron in a background $SU(2)$ gauge field \cite{PhysRevLett.108.216802}. That is, it is of the form
\begin{equation}
	\label{eq: H_chiral}
	\begin{aligned}
	\HH&= \begin{pmatrix}
		0 & \DD^\da \\
		\DD & 0
	\end{pmatrix} \\
	\DD &=2iv_0\qty(\bar{\partial}+\bar{A})
\end{aligned}
\end{equation}
where $\bar{\partial} = \frac{1}{2}(\partial_x+i\partial_y),\bar{A}=A_x-iA_y$ with $\vec{A}$ being a non-abelian traceless gauge potential (here we focus mostly on the $SU(2)$ gauge group). We assume that $\bar{A}$ is periodic on some lattice. The Hamiltonian $\HH$ comes naturally with a chiral symmetry, which we choose to be Hermitian, $S=\s_z$. We follow our notation for TBG and use $\s_i$ as the isospin indices and $\h_i$ as the gauge indices. 
For any two solutions $\y_{a,b}$ of $\DD \y=0$ (that is, Dirac cone wavefunctions in the same $S$ indices) we can define the Wronskian for the functions $\y_{a,b}$ as
\begin{equation}
	\label{eq: wronskian_requirement}
	\II\rarg=\y_{a,1} \rarg\y_{b,2}\rarg-\y_{b,1}\rarg\y_{a,2}\rarg,
\end{equation}
where the second index is a spinor index. In \cite{PhysRevB.103.155150} the authors show that $\II\rarg$ is position independent and further show that the condition for exactly flat bands is the existence of two orthogonal $\y_{a,b}$ for which $\II\rarg =0$ (see a review in Appendix \ref{appendix:wronskian-summary}). This condition may be expressed as 
\begin{equation}
	\ip{\mathcal{W}\y_a}{\y_b}=0,
	\label{Wronskian-condition}
\end{equation}
where $\mathcal{W}$ is the ``Wronskian operator" defined by 
\begin{equation}
	\label{wronskian-definition}
	\mathcal{W}=i\h_y K.
\end{equation}
The Wronskian is antiunitary, commutes with $S$, and satisfies $\mathcal{W}^2=-1$ (the choice $\WW=i\h_y\s_z K$ satisfies the same requirements and gives an equivalent condition). Since the Wronskian inverts the direction of the spinor that it operates on, Eq. (\ref{Wronskian-condition}) implies that 
\begin{equation}
    \psi_b=\psi_a\nu({\bf r}),
    \label{proportionality}
\end{equation} with $\nu({\bf r})$ being a scalar. 

Let us consider the symmetry groups which allow a Dirac Hamiltonian as in \eqref{eq: H_chiral}. We first assume that the symmetries of the system keep $\y_{a,b}$ orthogonal (this assumption is broken, e.g., in $C_3$-broken cTBG, which we treat in Appendix \ref{app: C_3 breaking}). Since we want the momentum operator to be diagonal in $\DD$, it must satisfy the condition
\begin{equation}
    \label{eq:dirac_matrices_relation}
	v_x v_y=i v_0^2 S.
\end{equation}
If the Hamiltonian has additional time-reversal symmetry $T$, then $T$ and $P=ST$ preserve the space of the two degenerate Dirac cones and therefore satisfy the definition \eqref{eq: TPS_commutators}. Using \eqref{eq: TPS_commutators} and \eqref{eq:dirac_matrices_relation} we then find that
\begin{align}
\label{eq: xit_xip}
	\qty{T,S}&=0& \Leftrightarrow&\,& T^2&=-P^2.
\end{align}
This requirement already restricts the possible AZ symmetry classes which can support a continuum Hamiltonian with exactly flat bands to AIII ($S$ only), DIII (where $P^2=-T^2=1$), and CI (where $T^2=-P^2=1)$. Notice that the Hamiltonian \eqref{eq: H_chiral}  can be thought of as a surface Hamiltonian for class AIII, CI, or DIII topological superconductors, all of which can have protected Dirac cones on the surface \cite{schnyder2008classification, schnyder2009lattice}. This proves that the Hamiltonian \eqref{eq: H_chiral} cannot open a gap at zero energy.

We now treat each one of the above symmetry classes. We first consider the two time-reversal symmetric classes, DIII and CI. We find that class DIII Hamiltonians are too constrained to allow for the condition \eqref{eq: wronskian_requirement}, while for class CI the condition can be fulfilled and is, in fact, equivalent to the vanishing of the Dirac velocity. We then show that the analysis of class AIII Hamiltonians can be mapped onto the analysis of class CI. Therefore, understanding the criteria for obtaining a flat band in the CI case is sufficient for the more general case of Hamiltonians of the form \eqref{eq: H_chiral} with no time-reversal symmetry.

\subsection{Class DIII}
Here we prove that a $4\times4$ Hamiltonian of the form \eqref{eq: H_chiral} of class DIII cannot support exactly flat bands. We begin by fixing $v_x=\s_x,v_y=\s_y,S=\s_z$ and $T=\s_y K$ (the choices $T=\h_x\s_y K$ and $T=\h_z\s_y K$, where $\h$ are the gauge field indices, are equivalent). This restricts $\bar{A}$ to be of the form
\begin{equation}
	\begin{pmatrix}
	0 & \bar{A}^\dagger\\
	\bar{A} & 0
	\end{pmatrix}=a_x(\bm r) \sigma_x\h_y+a_y\rarg\sigma_y \h _y
\end{equation} 
for some real $a_x\rarg,a_y\rarg$.  Equivalently, we can write
\begin{equation}
\label{eq:A_bar_diii}
    \bar{A}=
    \begin{pmatrix}
    0 & -i\\
    i & 0
    \end{pmatrix}a\rarg
\end{equation}
where $a\rarg=a_x\rarg+ia_y\rarg$. Since $\bar{A}$ commutes with itself at different positions the equation $\DD\y=0$ can be straightforwardly solved by integrating both sides. To do so we decompose $a\rarg$ as
\begin{equation}
	a(\bm r) =\sum_{\bm G} a_{\bm G} e^{\frac{1}{2}(z \bar{G}+\bar{z}G)}
\end{equation}
where the sum over $\bm G$ is a sum over reciprocal lattice vectors and $G=G_x+i G_y$.
The zero modes of $\DD$ can then be calculated explicitly as
\begin{align}
\label{eq:diii_zero_modes}
	\y_\pm(\bm r) &= \begin{pmatrix}
		1 \\ \pm i \end{pmatrix} f(z)e^{\pm u\rarg} \\
		\begin{split}
		    u\rarg & = \sum_{{\bm G}\neq0} \frac{2 a_{\bm G}}{G} e^{\frac{1}{2}(z \bar{G}+\bar{z}G)} \\
		    &\qquad+ \Re(a_0)(\bar{z}-z)+\Im(a_0)(\bar{z}+z)
		\end{split}
\end{align}
where $f(z)$ is holomorphic. Since $e^{\pm u(\bm r)}$ is periodic (up to a phase) and therefore bounded, $\y_\pm$ is normalizable only for $f(z)=\rm{const.}$ We can conclude that for any $a\rarg$ there are only  two zero modes for $\DD$, which are given by (\ref{eq:diii_zero_modes}). Since these solutions do not satisfy (\ref{proportionality}),  there are no exactly-flat bands for any choice of $a\rarg$. One can explicitly check that the Wronskian $\II\rarg$ of $\y_+$ and $\y_-$ is constant and nowhere vanishes, since the spinors are never parallel. 
\subsection{Class CI}
While class DIII symmetries limit $\bar{A}$ to the form \eqref{eq:A_bar_diii}, for class CI $\bar{A}$ does not necessarily commutes with itself at different points. As a result, the zero modes cannot be obtained by an integration procedure similar to \eqref{eq:diii_zero_modes}. This allows for a richer structure of the zero modes and, most interestingly to us, allows for the vanishing of $\II\rarg$.

Further notice that, for class CI Hamiltonians, the combination $v_xT$ satisfies
\begin{align}
	\qty(v_xT)^2 &=-v_0^2,\\
	\qty[v_xT,S] & = 0
\end{align}
and thus must be proportional to either $\h_yK$ or $\s_z\h_yK$. We therefore have
\begin{equation}
    \ip{\WW \y_a}{\y_b}= {\rm const. }\times \mel{T\y_a}{v_x}{\y_b}.
\end{equation}
Since the RHS is an element of $\r(v_x)$ the vanishing of the Dirac velocity implies \eqref{Wronskian-condition}. This argument shows that in class CI Hamiltonians, as long as the solutions $\y_a$ and $\y_b$ are kept orthogonal, the vanishing of the Dirac velocity implies the existence of exactly-flat bands. From Table \ref{tab:Double DC} we see that for class CI (that is, with $\Theta^2=+1,\Pi^2=-1$) we have $\zvc\le2$. 

\subsection{Class AIII}
For class AIII we can write $\bar{A}$ in the general form
\begin{equation}
    \bar{A}\rarg=\begin{pmatrix}
    W\rarg +Z\rarg & X\rarg+iY\rarg \\
    X\rarg-iY\rarg & W\rarg -Z\rarg
    \end{pmatrix}.
\end{equation}
Here $W$ represents the $U(1)$ part of the gauge potential (physically, $\nabla\times W$ is a magnetic field), while $X,Y,Z$ are the three components of the $SU(2)$ part. When $W=0$ this is a system of class CI with $T=\h_y\s_yK$. Let us first assume for simplicity that $W\rarg$ is periodic with mean zero (this represents a staggered magnetic field). We notice that, for a zero mode $\y$ of $\DD$, $W\rarg$ can be absorbed to $\y$ by defining
\begin{equation}
    \begin{aligned}
    \y'\rarg&=e^{\pbar^{-1}W\rarg}\y\rarg \\
    \bar{A}'\rarg&= \bar{A}\rarg-\mathbb{I}\cdot W\rarg\\
    \DD'&=2iv_0(\pbar+\bar{A}')
\end{aligned}
\end{equation}
where the operator $\pbar^{-1}$ is formally defined by
\begin{equation}
\label{eq:inv_pbar}
    \bar{\partial}^{-1} \qty(e^{-i \bm q\cdot\bm r})=\frac{2i}{q}e^{-i \bm q\cdot\bm r}
\end{equation}
with $q=q_x+iq_y$. Under this definition, we find that $\DD\y=0$ if and only if $\DD'\y'=0$. That is, we can reduce the problem of finding a zero mode for $\DD$ to the case in which $\bar{A}$ is traceless. We conclude that for finding exactly-flat bands of \eqref{eq: H_chiral} it is sufficient to solve for the time-reversal symmetric (CI) case, where, as we showed above, the vanishing of the Dirac velocity implies an exactly-flat band. 

The flat bands created by this procedure have an exact correspondence with lowest-Landau-level (LLL) wavefunctions \cite{wang2021exact, PhysRevB.103.155150, Sheffer2021}, and therefore have a nonzero Chern number for each $S$ polarization (they can, in fact, be written in a form that resembles the LLL wavefunctions on the plane, see Appendix \ref{appendix:wronskian-summary}). By considering the ``squared Hamiltonian" $\bar{H}=\HH^2$, for which $S$ acts as a local unitary symmetry, we see that the nonzero Chern number on each $S$ index gives the middle bands a $\Z\times\Z$ topological index when $T$ is absent. This index collapses to a (nonzero) $\Z$ index in the presence of $T$ \footnote{The transition at one flux quantum in magic-angle cTBG in a magnetic field \cite{PhysRevB.103.155150, Sheffer2021} can be seen as a transition from a (1,-1) index to (2,0).}. Finally, note that cTBG is in class CI, as a result of an emergent intra-valley $T$ symmetry \cite{Wang2021Chiral}.

It is important to distinguish between the exactly-flat-bands models discussed here and the flat bands in tight binding models, e.g. in bipartite lattices \cite{sutherland1986localization, bergman2008band, hwang2021general, cualuguaru2022general} or in line-graph lattices \cite{kollar2020line, PhysRevResearch.2.043414}. The models we discuss allow for the creation of exactly-flat bands by the tuning of a small number of parameters, assuming that a given set of symmetries are preserved. The small value of the codimension $\zvc$ implies that even when symmetry-allowed terms give the flat band a dispersion, this dispersion can always be compensated by the lowest-momentum tunneling, and the flatness is recovered. This property is not there in the tight-binding examples. The bipartite lattice models have a flat band in all possible parameters, provided that the lattice remains bipartite. For the line-graph lattices, on the other hand, there is an infinite set of parameters that may be varied to destroy the band flatness while preserving the lattice symmetries, and can not be compensated by other parameters. A further difference is that the models discussed here are continuum, rather than tight binding,  models. This property allows for the separation of the exactly-flat bands to bands of opposite Chern numbers by a symmetry-breaking perturbation, such as a sublattice potential in the case of cTBG. In lattice models, on the other hand, the existence of exactly-flat bands with nonzero Chern number is prohibited \cite{jian2013momentum, chen2014impossibility} (but such bands may carry fragile topology \cite{cualuguaru2022general}).

\section{Examples of class CI flat-band models}
\label{sec: applications of flat band models}
\subsection{Chiral $C_4$-symmetric model}
\begin{figure}[t]
	\centering
	\includegraphics[scale=.58]{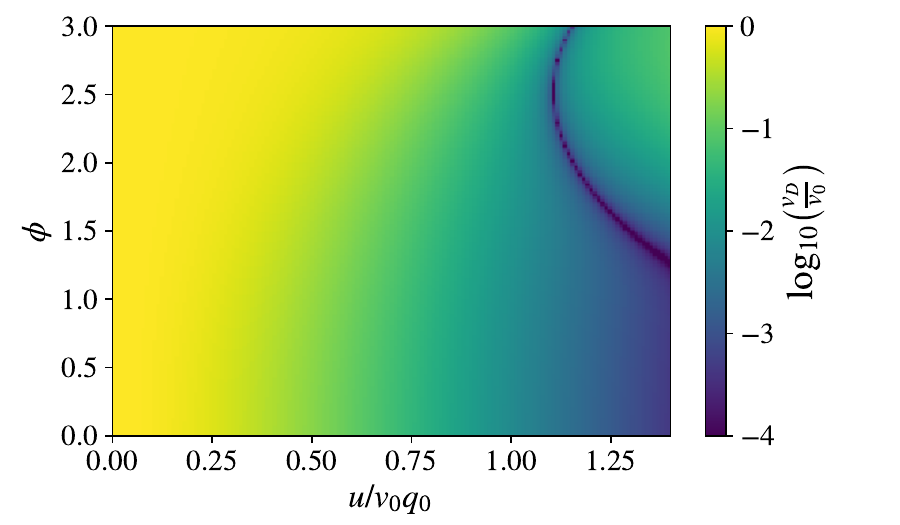}
	\caption{\label{fig:vanishing dirac velocity} Dirac velocity for the Dirac cone of the Hamiltonian \eqref{eq:H_mag_rashba_full} as a function of the parameters $u,\f$, in logarithmic scale. The Dirac velocity vanishes exactly on the dark line, leading to an exact flattening of the bands.}
\end{figure}
\begin{figure}[tb]
	\centering
	\includegraphics[scale=0.62]{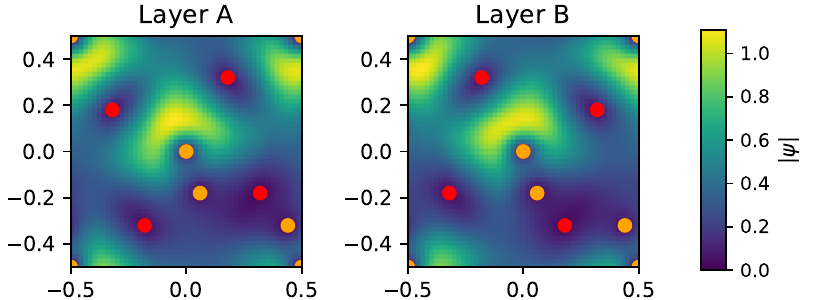}
	\caption{\label{fig:c4_wavefunction} Example wavefunction $\y_{\bm k=0}$ obtained numerically from the Hamiltonian \eqref{eq:H_mag_rashba_full} with $u=1.1,\f=2.5$ (at the flat band). Each layer has four mutual zeros, which are zeros of the entire wavefunction (in orange), and four zeros of opposite chirality which cancel the complex winding of the common zeros (in red).}
\end{figure}
The insights gained in the previous section can be used to construct a continuum Hamiltonian of class CI that can be tuned to have exactly flat bands at zero energy. Our model is manifestly distinct from cTBG in that it has a $C_4$, instead of $C_3$ symmetry. We will therefore call it the $C_4$ symmetric flat-band model (C4FB). The presence of the $C_4$ symmetry has additional interesting implications, which will be discussed shortly. 

The C4FB model consists of two Dirac cones on two topological insulator (TI) surfaces on the $x-y$ plane, connected by $z$-reflection symmetry, and coupled via a spin-dependent tunneling term modulated by an in-plane magnetic field (a system with slightly similar features was analyzed in \cite{chaudhary2022twisted}). The two 3D TI surface-states
are described by \cite{fu2007topological, moore2007topological, roy2009, hasan2010colloquium}:
\begin{equation}
	\HH_{\rm TI} = \h_z v_0 \bm p \cdot \bm \s
\end{equation}
where $\bm \s = (\s_x,\s_y)$. Each of these Dirac cones has a chiral symmetry $S=\sigma_z$ and a spinful time-reversal symmetry $T=K\sigma_y$, with $T^2=-1$. Since we want the system to be in class CI, we need to preserve $S$ and replace $T$ by a symmetry $T'$ satisfying $T'^2=+1$ and $\qty{T',S}=0$.  To that end, we introduce spin-flipping tunneling between the layers, whose phase is modulated by an in-plane magnetic field: 

\begin{equation}
	\begin{aligned}
	\label{eq:H_mag_rash}
	\HH_{\rm tunneling} = &\qty(\h_y \cos{A_z\rarg}+\h_x \sin{A_z\rarg})\qty(\bm u\rarg\cdot\bm \s)\\
	\end{aligned}
\end{equation}
where $A_z\rarg$ is the vector potential associated with the magnetic field. The resulting symmetry $T'=K\eta_x\sigma_x$ is a combination of $T$ and the $z$-reflection $R_z=\s_z\h_x$. The Hamiltonian is then
\begin{equation}
	\label{eq:H_mag_rashba_full}
	\HH=\HH_{\rm TI}+\HH_{\rm tunneling}.
\end{equation}

We additionally require the symmetries $C_4=e^{i\frac{\p}{4}\s_z}\qty({\bm r}\to\mathcal{R}_4{\bm r}),M_x=\s_y(x\to-x)$ and a translation invariance by the unit cell. While any form of $A_z$ and $\bm u$ satisfying these requirements will give similar results, we choose for concreteness
\begin{equation}
	\label{eq:H_parameters}
	\begin{aligned}
		\bm u\rarg &= u \qty(\sin \frac{2\p x}{a_0},\sin \frac{2\p y}{a_0}), \\
		A_z\rarg &= \f\qty(\cos \frac{2\p x}{a_0}+\cos \frac{2\p y}{a_0}).
	\end{aligned}
\end{equation}
Notice that $\HH$ can be made of the form \eqref{eq: H_chiral} by a gauge transformation.

By diagonalizing $\HH$ we find that, as expected from our calculation of $\zvc$ (we have $\Theta=T',\Sigma=S,R=M_x$, see table \ref{tab:Double DC}), the Dirac velocity vanishes on a codimension-one manifold in the $u,\f$ space (see Fig. \ref{fig:vanishing dirac velocity}). Our discussion above shows that when the Dirac velocity vanishes the two degenerate $\s_z=+1$ wavefunctions $\y_{a,b}$ at ${\bm k}=0$ satisfy $\II\rarg=0$. We can therefore write 
\begin{equation}
	\y_a\rarg=\nu(z)\y_b(\bm r)
\end{equation}
where $\pbar\nu=0$ since both $\y_{a,b}$ satisfy $\DD \y=0$. The function $\nu(z)$ is periodic on the lattice and $C_4$ symmetric, inherited from $\y_{a,b}$. Therefore $\nu\left(z\right)$ must have at least four poles per unit cell, located at four $C_{4}$-related points. At these four points $\y_{1}$ must be zero \footnote{A similar argument can be used in the cTBG Hamiltonian \eqref{eq: H_chiral_app} to find that there must be at least three zeros per unit cell for the translation symmetry of the model. Notice, however, that the Hamiltonian presented in the gauge choice of \eqref{eq: H_chiral_app} has a unit cell which is three times larger than the physical unit cell.}. Using the fact that $\y_{1}$ has four zeros per unit cell we can construct four $\s_z=+1$ linearly-independent zero-energy wavefunctions at each $\bm k$, in the form \cite{Ledwith2020}
\begin{align}
\label{eq:y_k_C4FB}
	\y_{\boldsymbol{k}}\left(\boldsymbol{r}\right) & =\Lambda_{\boldsymbol{k},n}\left(z\right)\y_{1}\left(\boldsymbol{r}\right),\\
	\Lambda_{\boldsymbol{k},n}\left(z\right) & =e^{i k_x z}\prod_{i=1,...,4}\frac{\vartheta_{1}\left(\frac{z-w_i}{a_0}\mid i\right)}{\vartheta_{1}\left(\frac{z-z_i}{a_0}\mid i\right)}.
\end{align}
where $\vartheta_{1}\left(z\mid\t\right)$ is the Jacobi theta function \cite{Whittaker1996} (we use the convention defined in \eqref{vartheta}),
$z_{i}=x_i+iy_i$ are the zeros of $\y_{1}$ and $w_{i}$ satisfy
\begin{equation}
	\label{eq:ki}
	a_0 k=\frac{2\p}{a_0}\sum_{i}w_{i}+n+mi;\ m,n\in \mathbb{Z}.
\end{equation}
where $k=k_x+ik_y$. That is, the positions of $w_{i}$ determine the momentum of $\y_{\boldsymbol{k}}$, and the possible configurations of $w_{i}$ satisfying (\ref{eq:ki}) give the four degenerate wavefunctions. This construction is similar to that of lowest-Landau-level wavefunctions on the torus \cite{Haldane1985}. Note that our arguments did not rule out the possibility of having more than four wavefunctions per unit cell, but we expect four to be the general case. We give an example of $\y_{\bm k=0}$ in Fig. \ref{fig:c4_wavefunction}. 

\subsection{Magic parameters in Hamiltonians with symmetry-protected quadratic band-touching}
Here we consider a system, previously analyzed in \cite{PhysRevResearch.4.043151}, which displays symmetry-protected quadratic band-touching (QBT). We show how the symmetry analysis can provide the condition in which a perfectly-flat band can be created. 

The Hamiltonian we consider will be of the form:
\begin{align}
    \label{eq: quadratic band Hamiltonian}
    \HH &= \begin{pmatrix}
        0 & \DD^*\rarg \\
        \DD\rarg & 0
    \end{pmatrix} \\
    \DD\rarg &= \frac{1}{2m_0}\pbar^2+u\rarg
\end{align}
where $u=u_x+iu_y$. In \cite{PhysRevResearch.4.043151} the authors suggest realizing this Hamiltonian by stacking two twisted layers of a material hosting a QBT point at the $K$-points. The Hamiltonian shown there realizes two copies of $\HH$ with 
\begin{equation}
\label{eq: u twisted quadratic}
u\rarg=u_x\rarg=\a\qty(\cos qx-\cos qy). 
\end{equation}
Below we provide a proof that $\HH$ with $u\rarg$ given by \eqref{eq: u twisted quadratic} has flat bands with codimension 1 in $\a$ (this was shown numerically in \cite{PhysRevResearch.4.043151}). \\

\subsubsection{From a QBT to a Dirac Hamiltonian}
As a first step, we show the relation between $\HH$ and the Hamiltonians discussed in Section \ref{sec: Exactly-flat bands}. We first note that $\HH$ has a time-reversal symmetry $\TT=\s_xK$ and a chiral symmetry $\SS=\s_z$, and is therefore in class CI, just as the flat-bands Hamiltonian discussed there. Furthermore, we can relate the 2-band Hamiltonian \eqref{eq: quadratic band Hamiltonian}, which has second-derivative operators, to a 4-band class CI Hamiltonian of the form \eqref{eq: H_chiral} having only first derivative operators, and having the same number of zero-energy states. 

To do so, we notice that for any scalar wavefunction $\y$ satisfying $\DD\y=0$, we have
\begin{equation}
    \tilde{\DD}\begin{pmatrix}\y \\ v_0\pbar\y\end{pmatrix} = 0 ,
\end{equation}
where
\begin{equation}
    \tilde{\DD}=\begin{pmatrix}
    v_0\pbar & -1 \\
    u & v_0\pbar
    \end{pmatrix}
\end{equation}
and $v_0=\qty(2m_0)^{-\frac{1}{2}}$. As a result, the four-component spinor $\qty(0,0,\y,v_0\pbar\y)$ is a zero-energy state of the Hamiltonian
\begin{equation}
    \tilde{\HH}=\begin{pmatrix}
        0 & \tilde{\DD}^\da \\ \tilde{\DD} & 0
    \end{pmatrix}.
\end{equation}

Notably, $\tilde{\HH}$ inherits the class CI symmetries of $\HH$, which are given by $\tilde{\TT}=\h_y\s_yK,\tilde{\SS}=\s_z$ ($\s_i,\h_i$ are the Pauli matrices in the spinor and gauge components, respectively). Similarly, it inherits any crystalline symmetry that $\HH$ has. This shows that finding conditions for a flat band in Hamiltonians of the form \eqref{eq: quadratic band Hamiltonian} is equivalent to finding the conditions for a flat band in class CI Dirac Hamiltonians, which were analyzed in Section \ref{sec: Exactly-flat bands}. There we showed that a flat band will be created as a result of a vanishing Dirac velocity, provided that $\tilde{\DD}$ has two orthogonal zero modes. We notice that here, for $u=0$, $\tilde{\HH}$ has a QBT, which amounts to having a vanishing Dirac velocity (i.e. a vanishing expectation of the operator \eqref{eq: velocity operator}) with only a single zero-mode of $\tilde{\DD}$. To conclude our argument, we show that when the quadratic term in the dispersion is made to vanish by the application of $u$, the QBT can be separated into two Dirac cones with vanishing velocity, resulting in perfectly-flat bands.

\subsubsection{From a Dirac Hamiltonian to a perfectly-flat band}
We now apply our analysis to provide conditions for the emergence of exactly-flat bands in $\HH$. We first notice that the QBT is stable when $u$ is modified if and only if $\tilde{\HH}$ is $C_4$ symmetric. When the QBT is unstable we get a similar situation to the one discussed in Appendix \ref{app: C_3 breaking}: the model has two zero-energy Dirac points, whose momenta change as $u$ is modified. As a result, the vanishing of the Dirac velocity will result in the Dirac points fusing to a QBT with nonzero quadratic dispersion, and exactly-flat bands will not generally form. To go on further we, therefore, need to assume that $\HH$ is $C_4$ symmetric with $C_4=\s_z \qty(x\to y,y\to -x)$ (which is also the case for the model discussed in \cite{PhysRevResearch.4.043151}).

With the assumption of a $C_4$ symmetry, we investigate which additional symmetries can help us make the bands perfectly flat. The form of $\tilde{\HH}$ ensures that the dispersion remains quadratic around the band-touching point, so the projected Hamiltonian near the $K$ point is restricted to the form
\begin{equation}
\begin{aligned}
    \tilde{\HH}_{\rm proj.}\qty(\bm k) &= \begin{pmatrix}
        0 & \qty(f_1(\a)+i f_2(\a))k^2 \\ \qty(f_1(\a)-if_2(\a))\bar{k}^2 & 0
    \end{pmatrix}\\ &+ O(k^4).
    \end{aligned}
\end{equation}
Consequently, the quadratic dispersion of $\tilde{\HH}$ vanishes with codimension 2. To reduce the codimension to 1 we require an additional symmetry that guarantees the vanishing of $f_2$, which is a reflection symmetry. Acting on $\HH$, this additional symmetry is equivalent to the requirement that $u$ is either purely real or purely imaginary.

Having shown that the quadratic term in the dispersion can be made to vanish, we now show that this vanishing leads to perfectly-flat bands. To do so, we map the problem to the problem studied previously, where we had two separate zero modes $\y_{a,b}$ of $\tilde \DD$. This can be done by adding a small $C_4$-breaking perturbation to $u$, of the form
\begin{equation}
    u\to u-v_r(\a) k_\e
\end{equation}
where $v_r(\a)=\sqrt{f_1(\a)+if_2(\a)}$ is the square-root of the renormalized quadratic dispersion around the $K$ point and $k_\e$ is small. The resulting Hamiltonian has, therefore, two Dirac points at $\bm k=(\pm k_\e,0)$, which remain separate for any $\a$. In addition, the Dirac velocity of the Dirac cones vanishes whenever $\a$ is tuned to make $v_r$ vanish. From the analysis of section \ref{sec: Exactly-flat bands}, we conclude that $\tilde{\HH}$ with the additional perturbation has exactly-flat bands whenever $v_r$ vanishes and, taking $k_\e$ to zero, we conclude that $\tilde{\HH}$ and therefore $\HH$, has exactly-flat bands whenever $v_r$ vanishes. 

The analysis in this section shows, therefore, that Hamiltonians of the form \eqref{eq: quadratic band Hamiltonian} with a $C_4$ symmetry have exactly-flat bands with a small codimension (1 if there is a reflection symmetry present and 2 otherwise). We note that, while an example of such a Hamiltonian was first provided in \cite{PhysRevResearch.4.043151} using two twisted layers of a 2D material with QBT points, the $u$-term in \eqref{eq: quadratic band Hamiltonian} can be obtained by a long-wavelength modification of the hopping in a single layer hosting a QBT point (for example by periodic strain from surface-acoustic waves). 

\subsection{Quasi-crystalline generalization of cTBG}
\begin{figure*}
    \centering
    \includegraphics[scale=.7]{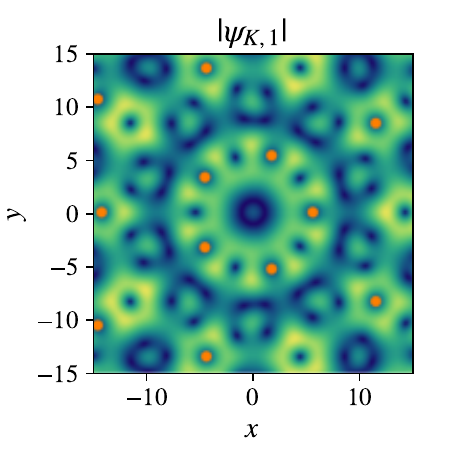}
    \includegraphics[scale=.7]{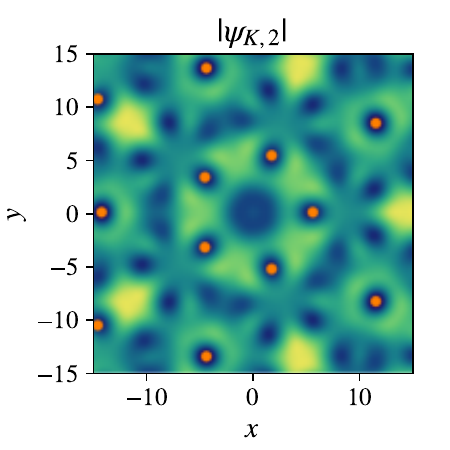}
    \caption{The wavefunction $\y_K$ in the chiral $C_5$ symmetric model \eqref{eq:c_n_matbg} at the magic angle. The orange points signify zeros of the wavefunction.}
    \label{fig:C_5_psi}
\end{figure*}
Here we discuss a quasi-crystalline generalization of the cTBG Hamiltonian, namely the generalization of the $C_3$-symmetric model to a $C_n$ symmetric model for odd $n\ge3$. We focus on the chiral case as it is the easiest to analyze theoretically. The family of Hamiltonians is given by the form \eqref{eq: H_chiral} with
\begin{equation}
\begin{aligned}
\label{eq:c_n_matbg}
    \bar{A} &=\begin{pmatrix}
    0 & \frac{\a}{2} U\rarg \\ \frac{\a}{2} U(-\bm r) & 0
    \end{pmatrix},\\
    U\rarg &= \sum_{j=0}^{n-1}e^{i\frac{2\p}{n}j}e^{-i\bm q_i\cdot\bm r}
    \end{aligned}
\end{equation}
with $\bm q_j=\qty(\cos(2\p j/n-\p/2)),\sin(2\p j/n-\p/2))$. Clearly \eqref{eq:c_n_matbg} reduces to the cTBG Hamiltonian \cite{Tarnopolsky2019} for $n=3$. \\
For $n>3$ the model is not crystalline anymore, but nevertheless the formal analysis of magic angles in cTBG continues to hold. That is, we can calculate by perturbation theory in $\a$ the correction to the zero-energy wavefunctions. We get a zero-energy wavefunction of the form
\begin{equation}
\label{eq: qc wavefun}
    \y_K\rarg=\sum_{n=0}^\infty \qty( \a\bar{\partial}^{-1}\bar{A})^n\begin{pmatrix}
    1 \\ 0
    \end{pmatrix}
\end{equation}
where the operator $\bar{\partial}^{-1}$ is defined in \eqref{eq:inv_pbar}. Note that $\bar{\partial}^{-1}$ is undefined for $\bm q=0$, but the $C_n$-symmetry prevents zero-momentum terms from appearing in the perturbation series. Another zero-energy wavefunction is given by acting on $\y_K$ with the intra-valley $C_2$ symmetry $C_2=\h_y (\bm r\to\bm -r)$. 

While the Dirac velocity is no longer well-defined (as there are no Bloch wavefunctions), the Wronskian operator $\WW$ given by \eqref{wronskian-definition} still is. Since the ``Dirac cone" wavefunctions are still reflection symmetric, we have $\zvc=1$ for the vanishing of the formal Dirac velocity given by
\begin{equation}
\label{eq: Cn_vd}
    v_D=\mel{C_2 \y_K}{\WW}{\y_K}=\braket{\y_K(-\bm r)}{\y_K\rarg}.
\end{equation}
When $v_D$ vanishes we find from equation \eqref{nu-er} that $\y_K$ must have extensively-many zeros (that is, the number of zeros in a given area being proportional to the area. By repeating the analysis in Appendix \ref{appendix:wronskian-summary} we, therefore, find a ``band" with an extensive degeneracy of zero-energy wavefunctions. 

In Appendix \ref{app: c_n perturbation} we describe the results of a perturbative calculation of $v_D$ similar to the one detailed in \cite{Tarnopolsky2019}. For $n=5$ we find the first ``magic angle" at $\a_0=.32$. We plot the resulting wavefunction in Fig. \ref{fig:C_5_psi}. Each zero of the wavefunction can be used to construct a zero-energy state. If the system is confined to a finite size $L$, the number of zero energy states (up to corrections of order $1/L$), will equal the total number of zeros of the magic-angle wavefunction, and will be extensive with the system size.

\section{Discussion}
\label{sec: discussion}
In this work, we discussed the symmetry structure that is required for the flattening of bands of Dirac fermions. We started with the requirement for the Dirac velocity to vanish and gave examples in TBG and a Dirac cone on the surface of a 3D topological insulator. Afterward, we discussed the vanishing of any dispersion, namely the symmetry requirements for the formation of exactly flat bands. We showed that, for a certain set of symmetries, the vanishing of the Dirac velocity implies that the band is exactly flat. 

The symmetry considerations which allowed us to calculate $\zvc$ do not provide us with a recipe for writing a Hamiltonian with $\zvc$ parameters which can have a vanishing velocity, but generically suggest natural candidates for flat-band Hamiltonians. In one of the cases, which we studied in section \ref{sec: applications of zvc}, the existence of an extra symmetry seemed to impede such vanishing by the most natural candidate Hamiltonian. This observation may indicate that there may be further symmetry considerations that may guide the search for such Hamiltonians. These are left here as a subject for future research. 

While in this work we focused mainly on 2D moir\'e materials, much of our discussion can be straightforwardly extended to other systems and different tuning parameters. For example, one can consider 3D nodal line materials, where each $k_z$ slice can be viewed as a 2D subsystem, and $k_z$ can serve as an adiabatic parameter. Another interesting future question is the generalization of our results in section \ref{sec: Exactly-flat bands} to the case of $SU(N)$ gauge fields. Studies of specific models, such as alternating-twist n-layer graphene \cite{Khalaf2019} and chiral twisted graphene multilayers \cite{wang2022hierarchy,ledwith2022family} show the richness of states that might arise in such cases. In the former, one can encounter exactly-flat bands coexisting with dispersive bands, while in the latter we see exactly flat bands with Chren numbers $C>2$. It would be interesting to see whether it is possible to give a classification of the possible states in that case, in terms of the underlying symmetries.

Finally, another interesting direction is an experimental realization of the models we presented here. The flat-band models discussed in section \ref{sec: applications of flat band models} are theoretically intriguing, but more work is needed if one wishes to find candidates for experimental systems which host them. On the other hand, we believe that the TI models discussed in section \ref{sec: applications of zvc} can be realized using currently available experimental capabilities. Such an increase in the density of states on the surface of a TI could give rise to intrinsic superconductivity or correlated insulators \cite{guerci2022designer} on the surface of a TI, or, more exotically, a gapped state that is symmetric to both time-reversal and charge conservation. Such a state must be topologically ordered with quasiparticles satisfying non-abelian statistics \cite{PhysRevB.84.235145, bonderson2013time, PhysRevB.88.115137, PhysRevB.89.165132, PhysRevB.92.125111, Stern2013}. 

\begin{acknowledgements}
    We thank Ohad Antebi, Sebastian Huber, and B. Andrei Bernevig for enlightening discussions and Daniel Kaplan for reading an early version of the manuscript. A.S. and Y.S. acknowledge support from the Israeli Science Foundation Quantum Science and Technology grant no. 2074/19, the CRC 183 of the Deutsche Forschungsgemeinschaft. This project has received funding from the European Research Council (ERC) under the European Union’s Horizon 2020 research and innovation program (grant agreement No. 788715, Project LEGOTOP).
\end{acknowledgements}

\appendix
\section{Rigorous definition of $\zvc$}
\label{app:rigorous_zvc}
Here we provide a more rigorous notion of $\zvc$. Specifically, we prove the following theorem:\\

\textbf{Theorem:} Let $H_{\vec{\a}} \qty(\bm k)$ be a Bloch Hamiltonian with $n_D$ degenerate Dirac cones at $\bm k = \bm k_D$ with energy $E_D$ (in general both $k_D$ and $E_D$ can depend on $\vec{\a}$), such that $H_{\vec{\a}} \qty(\bm k_D)$ is symmetric under a group $G$. We assume  that $H_{\vec{\a}}$ is controlled by a set of continuous parameters $\vec{\a}=\a_1,...,\a_d$ such that the Dirac point has the same degeneracy for all values of $\vec{\a}$. Further assume that for some parameter choice $\vec{\a}_0$ the Dirac velocity matrices
\begin{equation}
    \label{eq:v_appen}
	\r\qty(v_i)_{mn} = \mel{\y_m}{v_i}{\y_n}
\end{equation}
vanish and that the gap between the degenerate Dirac cone wavefunctions and the higher bands is not closed. Then there exists (locally) a manifold of dimension $\ge d-\zvc$ in $\vec{\a}$ space in which \eqref{eq:v_appen} vanishes. Here $\zvc>0$ is defined by \eqref{eq:v_decomposition} as the dimension of the vector space $V$ of tuple of matrices $(M_x,M_y)$ satisfying \eqref{eq:rep_comm}. \\

Before proving the theorem, a few notes are in order: 
\begin{enumerate}
    \item The symmetries in $G$ can be either unitary or antiunitary. We also allow for symmetries that anticommute with $H\qty(\bm k_D)$.
    \item Here $\zvc$ is an \textit{upper bound} to the codimension of the zero-velocity manifold. Cases where $\zvc$ is strictly larger than the codimension should arise in the case where there are additional low-energy emergent symmetries at the Dirac cones. An example can be given in the $C_3$-broken cTBG Hamiltonian (eq. \eqref{eq:cMATBG} below): In this case, the exact Hamiltonian does not have a rotational symmetry relating $v_x$ and $v_y$. On the other hand, the velocity operators satisfy \eqref{eq:dirac_matrices_relation}, giving rise to an additional constraint on the codimension.
    \item When the gap with the upper bands closes the Dirac velocity representations are no longer required to be continuous since $\y_i$ are no longer continuous. A gap closing can therefore create a boundary (of dimension $<d-\zvc$) to the zero-velocity manifold. We give an example of this scenario in Appendix \ref{app: C_4_breaking}.
\end{enumerate}

\begin{proof}
Since the gap between the degenerate point and the other bands does not close, we can calculate the correction to $\r(\hat{o})$ for any operator $\hat{o}$ via first-order perturbation theory, that is
\begin{equation}
\begin{aligned}
    \pdv{\r(\hat{o})_{mn}}{\vec{\a}} &= \mel{\pdv{\y_m}{\vec{\a}}}{\hat{o}}{\y_n} + \mel{\y_m}{\pdv{\hat{o}}{\vec{\a}}}{\y_n} \\
    &+ \mel{\y_m}{\hat{o}}{\pdv{\y_n}{\vec{\a}}}
    \end{aligned}
\end{equation}
where for the velocity operator
\begin{align}
    \pdv{\y_n}{\vec{a}}&=\sum_i{\vphantom{\sum}}' \frac{\mel{\y_i}{\pdv{H}{\vec{\a}}}{\y_n}}{E_i-E_n}\ket{\y_i}\\
    \pdv{\hat{o}}{\vec{\a}}&= \pdv{H}{\vec{\a}}{k_i}
\end{align}
with the sum running only on $\y_i$ outside the degenerate space. In the case of unitary operators $g\in G$ which preserve the degenerate subspace we have
\begin{align}
    \mel{\y_m}{g}{\pdv{\y_n}{\vec{\a}}}&=0\\
    \pdv{g}{\vec{\a}}&=0,
\end{align}
so we find $\partial_{\vec{\a}}\r(g)=0$. We then get for $v_i$ that
\begin{equation}
\label{eq:grad_rho_comm}
\begin{aligned}
    \r(g)^{-1}\pdv{\r(v_i)}{\vec{\a}}\r(g)
    &=\pdv{\vec{\a}}\qty(\r(g)^{-1}\r\qty(v_i)\r(g))\\
    &=\pdv{\vec{\a}}\r(g^{-1} v_i g).
    \end{aligned}
\end{equation}
We also have
\begin{equation}
\label{eq:g_comm_consts}
    g^{-1} v_i g=\sum_j \g_{i,j}^g v_j
\end{equation}
where $\g_{i,j}^g$ are real constants that depend on whether $g$ commutes or anticommutes with $H$, as well as the transformation that $g$ induces on $k$. Combining \eqref{eq:grad_rho_comm} and \eqref{eq:g_comm_consts} gives
\begin{equation}
\label{eq:grad_rho_consts}
    \r(g)^{-1} \pdv{\r(v_i)}{\vec{\a}} \r(g)=\sum_j \g_{i,j}^g \pdv{\r(v_j)}{\vec{\a}}.
\end{equation}
For a given set of matrix representations $\r(g)$ for $g\in G$ \eqref{eq:grad_rho_consts} gives a set of linear equation on the tuples $(\r(v_x),\r(v_y))$. The tuples satisfying \eqref{eq:grad_rho_consts} then form a vector space whose dimension is $\zvc$ (see the definition of $\zvc$ in \eqref{eq:v_decomposition})). There must therefore be at least $d-\zvc$ directions in $\vec{\a}$ space in which the velocity doesn't change, giving a (local) zero-velocity manifold around $\vec{\a}_0$ whose dimension is at least $d-\zvc$.
\end{proof}

\section{Review of the Bistritzer-Macdonald model and symmetries}
\label{app: MATBG}
Here we review the continuum model of twisted bilayer graphene (TBG) proposed by Bistritzer and Macdonald \cite{Bistritzer2011a}. 
\subsection{The TBG Hamiltonian}
The Bistritzer-Macdonald (BM) Hamiltonian  describes twisted bilayer graphene at small angles and low energies, at a single valley of the graphene layers. It is given by \cite{Bistritzer2011a, Bistritzer2011, ledwith2021strong}
\begin{align}
\label{eq:BM_H}
H & =\begin{pmatrix}h\left(-\q/2\right) & T\left(\bm{r}\right)\\
T^\dagger\left(\bm{r}\right) & h\left(\q/2\right)
\end{pmatrix},\\
h(\q) & =-iv\bm{\s}_{\q}\cdot \bm{\na},\\
T\left(\bm{r}\right) & =w\sum_{j}e^{-i\bm{q}_{j}\cdot\bm{r}}T_{j},
\end{align}
where $\bm{\s}_{\q}=e^{-i\q\s_{z}/2}\left(\s_{x},\s_{y}\right)e^{i\q\s_{z}/2}$. The single-layer $h$ are the Hamiltonians for a single Dirac cone in each graphene layer, twisted by a small angle. The tunneling matrices $T_{i}$ are
\begin{equation}
\begin{aligned}
T_{1} & =\begin{pmatrix}\k & 1\\
1 & \k
\end{pmatrix}, \\
T_{2,3} & =\begin{pmatrix}\k & e^{\mp i\f}\\
e^{\pm i\f} & \k
\end{pmatrix},
\end{aligned}
\end{equation}
with $\f=2\p/3$ and $\bm{q}_{1}=k_{\q}\left(0,-1\right),\bm{q}_{2,3}=k_{\q}\left(\pm\sqrt{3},1\right)/2$.
We have $k_{\q}=2\sin\left(\q/2\right)k_{D}\approx\q k_{D}$ where
$k_{D}=\frac{4\p}{3\sqrt{3}a_{0}}$ and $a_{0}\approx1.4$\ring{A}
is the distance between atoms in graphene. The scale $w\approx 110\si{meV}$ is the energy scale associated with the tunneling between the layers and the factor $0\le\kappa\le1$ determines the ratio between $AA$ and $AB$ tunneling between the sublattices of the graphene layers. Real-world TBG has $\kappa\approx 0.7$ as a result of lattice relaxation \cite{PhysRevB.96.075311}. 

Important to some of our discussion is the chiral limit of TBG (cTBG) obtained by setting $\k=0$. Under this assumption, we can remove the $\theta$ dependence in $h(\q/2)$ by a gauge transformation. To write the resulting Hamiltonian in a form compatible with \eqref{eq: H_chiral} we further re-scale the Hamiltonian by defining $\HH=H/E_0$ where $E_0=k_\q w$; define the dimensionless parameter $\a = w/k_\q v$; and rearrange the rows so that the Hamiltonian acts on the spinor $(\y_1,\y_2,\chi_1,\chi_2)$ (here the indices are layer indices, and $\y,\chi$ live on the $A,B$ sublattices, respectively). We obtain the chiral Hamiltonian \cite{PhysRevLett.108.216802,Tarnopolsky2019}
\begin{equation}
\label{eq: H_chiral_app}
\begin{aligned} \HH_\mathrm{chiral} & =\begin{pmatrix}0 & \DD^{*}\left(-\bm{r}\right)\\
\DD\left(\bm{r}\right) & 0
\end{pmatrix},\\
\DD\left(\bm{r}\right) & =
\begin{pmatrix}
-2i k_\q^{-1}\bar{\partial} & \a U\left(\bm{r}\right)\\
\a U\left(-\bm{r}\right) & -2i k_\q^{-1}\bar{\partial}
\end{pmatrix},
\end{aligned}
\end{equation}
where $z=x+i y,\bar{\partial}=\frac{1}{2}(\partial_x+i\partial_y)$ and $U(\bm r)=e^{i\bm q_1\cdot\bm r}+e^{i\f}e^{-i\bm q_2\cdot\bm r}+e^{-i\f}e^{-i\bm q_2\cdot\bm r}$. 
\subsection{Symmetries}
Let us discuss the symmetries of the BM Hamiltonian \eqref{eq:BM_H}. We define the Pauli matrices $\s_i,\h_i$ in sublattice and layer space, respectively. The point symmetries acting within the valley are given by \cite{po2018origin}
\begin{equation}
    \begin{aligned}
    C_2T:& &\s_x K(\bm r\to-\bm r), \\
    C_3:& &e^{-i\frac{2\p}{3} \s_z}(\bm r\to R_3\bm r), \\
    C_{2,x}:& & \h_x\s_x (y\to -y),
    \end{aligned}
\end{equation}
where $K$ is the complex conjugation operator and $R_3$ is the rotation matrix by $2\p/3$. Of the three symmetries described above, only the first two preserve the Dirac points. \\
As a result of the small angle between the layers, the BM Hamiltonian has an additional approximate particle-hole symmetry. If we take the approximation of setting $\q=0$ in $h(\q)$ the resulting Hamiltonian has a particle-hole (PH) symmetry given by \cite{Song2019} 
\begin{equation}
    \begin{aligned}
    \mathcal{C}:& &\h_y \s_x K.
    \end{aligned}
\end{equation}
In the real-world model of TBG, the symmetry is broken in order $\mathcal{O}(\q)$. The combination $\mathcal{C}C_{2,x}$ gives an additional antisymmetry that preserves the Dirac cone, that is
\begin{equation}
    \begin{aligned}
    \mathcal{C}C_{2,x}:& &\h_z K (y\to-y).
    \end{aligned}
\end{equation}

Finally, the Chiral model has, besides $\mathcal{C}$, the additional chiral symmetry
\begin{equation}
\begin{aligned}
    S:& &\s_z.
\end{aligned}
\end{equation}
Since in the Chiral model the $\q$ dependence in $h$ is removed by a gauge transformation, the unitary PH symmetry $\mathcal{C}$ is exact here. We can therefore combine $S$ and $\CC$ to obtain an intra-valley unitary rotation symmetry that sends $\bm r\to-\bm r$ \cite{Wang2021Chiral}. By combining the intra-valley rotation with $C_2T$ we obtain the intra-valley time-reversal symmetry $\mathcal{T}'=\s_y\h_yK$ which satisfies $\qty(\mathcal{T}')^2=+1$. This shows that the cTBG model is indeed in class CI.

\section{Additional parameters for tuning a $C_2$-symmetric vanishing-velocity Dirac cone}
\begin{figure}
    \centering
    \includegraphics[scale=.85]{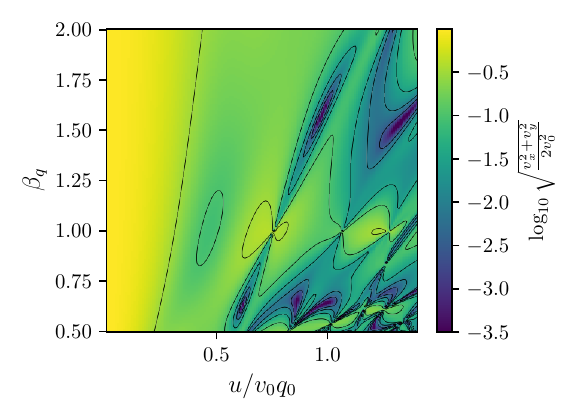}
    \caption{The ``absolute Dirac velocity" of the Dirac cone at charge neutrality of the Hamiltonian \eqref{eq: TI_hamiltonian_app} (similar to Fig \ref{fig:TI_vel_2d}a). The dark valleys are points of vanishing velocity.}
    \label{fig:variable_q_vel}
\end{figure}

Here we elaborate on our discussion of the Dirac cone on the surface of a 3D TI. In particular, we study additional parameters (besides the potential amplitude) which can be tuned to obtain a vanishing velocity for a  $C_2$-symmetric Dirac cone in a potential. The Hamiltonian of the form \eqref{eq: TI_hamiltonian} and \eqref{eq:TI_potential}
is defined to be consistent with the $T,M_x$, and $C_2$ symmetries, but is not the most general form consistent with these symmetries. More generally we can write an anisotropic form for the Dirac cone
\begin{equation}
    \HH=v_x\s_x p_x+v_y \s_y p_y + 2 u_x \cos q_x x + 2 u_y \cos q_y y.
\end{equation}
Here $v_x/v_y$ can be controlled by applying strain on the TI while $q_x/q_y$ can be controlled (for example) by an asymmetry in the  dielectric pattern. By rescaling the $y$ axis we can make $v_y=v_x=v_0$. We, therefore, write the Hamiltonian
\begin{equation}
\label{eq: TI_hamiltonian_app}
    \HH=v_0\bm p\cdot \bm \sigma + 2 u \qty( \cos q_0 x + \b_u \cos q_0\b_q y).
\end{equation}
which is controlled by the dimensionless parameters $\b_u,\b_q,u/q_0v_0$ (the first two define the $C_4$-symmetry breaking). In Fig. \ref{fig:variable_q_vel} we plot the velocity of the Dirac cone of \eqref{eq: TI_hamiltonian_app} at charge neutrality as a function of $u,\beta_q$. The results show similar ``magic parameters" to the case discussed in the main text (see Fig. \ref{fig:TI_vel_2d}a).
\section{The Wronskian operator and requirements for exactly flat bands}
\label{appendix:wronskian-summary}
In this section, we restate some results from \cite{PhysRevB.103.155150} that are useful to our discussion of the condition of exactly flat bands in chiral-symmetric continuum models. We begin with a Hamiltonian of the form
\begin{equation}
	\label{eq: H_chiral_app_general}
	\begin{aligned}
	\HH&= \begin{pmatrix}
		0 & \DD^\da \\
		\DD & 0
	\end{pmatrix} \\
	\DD &=-2iv_0\qty(\bar{\partial}+\bar{A})
\end{aligned}
\end{equation}
where $\bar{\partial} = \frac{1}{2}(\partial_x+i\partial_y), \bar{A}=A_x-iA_y$ with $\vec{A}$ being an $SU(2)$ gauge potential. We assume that $\HH$ is symmetric under translations by the lattice vectors $\bm a_1,\bm a_2$. Given two solutions $\y_a,\y_b$ of the zero mode equation
\begin{equation}
    \DD \y(\bm r)=0
\end{equation}
one can write the Wronskian
\begin{equation}
    \II\rarg=\y_{a,1} \rarg\y_{b,2}\rarg-\y_{b,1}\rarg\y_{a,2}\rarg.
\end{equation}
Importantly, we find that $\II\rarg=\rm const.$ This is because
\begin{align*}
    \pbar\II(\bm r)&=i\pbar\qty( \y_a^T\h_y\y_b)=-i\y_a^T\qty(\bar{A}^T\h_y+\h_y\bar{A})\y_b \\
    &= -i\y_a^T\qty(\h_y \tr\bar{A})\y_b=-\tr\bar{A}\cdot\II(\bm r).
\end{align*}
When there is no external magnetic field we have $\tr{\bar{A}}=0$, so $\II\rarg=\II(z)$. However, since $\II(z)$ has no singularity it must be constant. 

Next, we find that when $\II=0$ there must be an exactly flat band at zero energy. This is because, in that case, we must have
\begin{equation}
    \label{nu-er}
    \y_b(\bm r) = \nu\rarg\y_a(\bm r).
\end{equation}
We have, however, $\pbar\nu\rarg=0$, as can be seen by applying $\DD$ on both sides of \eqref{nu-er}. We can therefore write $\nu\rarg=\nu(z)$. Assuming that $\y_{a,b}$ are orthogonal (which must be the case if they are positioned on different points in the BZ) $\nu(z)$ is a non-constant meromorphic function. It therefore must have a pole at some point $z_0$ in the unit cell. At this point, $\y_a$ must have a zero. We can then construct additional wavefunctions in the flat band by writing
\begin{equation}
\label{eq:psik_generation}
    \y_{\bm k}\rarg=\frac{\vartheta_1\qty(\frac{z-z_0}{a_1}+\frac{\bm k\cdot \qty(\omega \bm a_1 -\bm a_2)}{2\p}\mid \frac{a_2}{a_1})}{\vartheta_1\qty(\frac{z-z_0}{a_1}\mid \frac{a_2}{a_1})}
                        e^{i\qty(\bm k\cdot \bm a_1)\frac{z}{a_1}}
                        \y_a(\bm r)
\end{equation}
where $a_i=\bm a_{i,x}+\bm a_{i,y}$, with $\bm a_i$ being lattice vectors. Here $\vartheta_1(z\mid\tau)$ is the Jacobi theta function, defined by
\begin{equation}
\label{vartheta}
\begin{aligned}
    \vartheta_1(z\mid\tau)&=\sum_{n=-\infty}^\infty (-1)^{n-1/2} e^{i\pi(n+1/2)^2\tau} e^{2\pi i(n+1/2)z} \\
        &=2\sum_{n=0}^\infty(-1)^n e^{i\p(n+1/2)^2 \tau}\sin{(2\p(n+1/2)z)}.
\end{aligned}
\end{equation}
Importantly, the pole of the $\vartheta_1$ cancels the zero at $\y_a$ making $\y_{\bm k}$ as defined above normalizable. 

One can follow an alternative approach for the construction of the flat-band wavefunctions, which makes the similarity between the flat-band wavefunctions and the lowest Landau levels manifest. We can choose a basis of such functions in the form \cite{Sheffer2021}
\begin{equation}
    \y\rarg=f(z)e^{-\frac{\p}{2A}\qty|z|^2}G\rarg
\end{equation}
where $f(z)$ is any holomorphic function, $A$ is the unit cell area, and $G\rarg$ is a structure function that captures the lattice dependency of the wavefunction. It is given by
\begin{equation}
    G\rarg=\frac{e^{\frac{\p}{2\Im (a_2/a_1)}\qty(\qty|\frac{z}{a_1}|^2+\qty(\frac{z}{a_1}-2i\Im\frac{z_0}{a1})^2)}}
    {\vartheta_1\qty(\frac{z-z_0}{a_1}\mid \frac{a_2}{a_1})}
                        \y_a(\bm r).
\end{equation}
Interestingly, it can be checked that $\qty|G\rarg|$ is periodic with the lattice.

\begin{figure*}
    \centering
    
    \begin{tikzpicture}
        \node at (-2,.5) {(a)};
        \node at (0,0) {\includegraphics[page=1,scale=1.15]{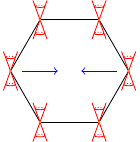}};
        \node at (3.75,0) {\includegraphics[page=2,scale=1.15]{c3_breaking_drawings.pdf}};
        \node at (0,-3) {\includegraphics[page=3,scale=1.15]{c3_breaking_drawings.pdf}};
        \node at (3.75,-3) {\includegraphics[page=4,scale=1.15]{c3_breaking_drawings.pdf}};
        
        \draw[double distance=1pt,->] ({3.75*(1/2)-.4},0) -- ({3.75*(1/2)+.4},0);
        \draw[double distance=1pt,->] ({3.75/2+.8},-.7) -- ({3.75/2-.8},-2.3);
        \draw[double distance=1pt,->] ({3.75*(1/2)-.4},-3) -- ({3.75*(1/2)+.4},-3);
    \end{tikzpicture} \qquad
    \includegraphics[scale=.7]{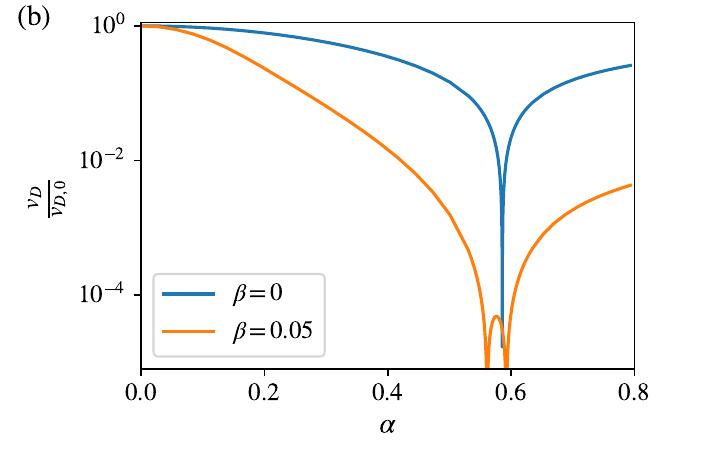}

    \caption{\label{fig:Dirac points trajectory} (a) The trajectory of the Dirac points in a $C_3$ broken cTBG near the magic angle when $\a$ is varied. For a small symmetry-breaking parameter $\b$ the Dirac cones remain close to the $K,K'$ points away from the magic angle. Near the magic angle, the displacement $\d \bm k_D$ diverges and the Dirac cones travel around the BZ, meeting twice to form QBT points at the $\Gamma$ and $M$-points.\\
    (b) The normalized Dirac velocity at the Dirac cones for $C_3$-symmetric ($\beta=0$) and weakly $C_3$ broken ($\beta=0.05$) cTBG Hamiltonian \eqref{eq:cMATBG}. In the $C_3$ broken case, the Dirac velocity vanishes twice, at the two QBT points.}
    \label{fig:my_label}
\end{figure*}
\section{$C_3$ symmetry breaking in cTBG}
\label{app: C_3 breaking}
Here we discuss the effects of $C_3$ symmetry breaking in cTBG. For concreteness, we consider the Hamiltonian \eqref{eq: H_chiral_app_general} with
\begin{equation}
    \begin{aligned}
        \label{eq:cMATBG}
        \DD &= \begin{pmatrix}
        -2ik_\q^{-1}\pbar & \a U\rarg \\
        \a U(-\bm r) & -2ik_\q^{-1}\pbar
        \end{pmatrix} \\
        U(\bm r) &= (1+\b) e^{-i \bm q_1\cdot \bm r}+e^{i\f}e^{-i \bm q_2\cdot \bm r} + 
        e^{-i\f}e^{-i \bm q_3\cdot \bm r} \\
    \end{aligned}
\end{equation}
where $\bm q_1 =k_\q (0,-1),\bm q_{2,3} = k_\q (\pm \sqrt{3}/2,1/2)$ and $\f=2\p/3$. Here $\a$ is the layer coupling scale and $\beta\ll 1$ is the $C_3$ symmetry-breaking scale. 

Since the Hamiltonian still has a chiral and a reflection symmetries, each of $v_x,v_y$ vanish with codimension 1 (as can be read from Table \ref{tab:single DC}). We also see that $\zvc=1$, since besides the symmetry requirements we have the additional relation between the chiral symmetry and the velocity operators, given by
\begin{equation}
\begin{aligned}
    v_y&=-iS v_x \\
    \Rightarrow  \r(v_y)&=-i\r(S) \r(v_x)
    \end{aligned}
\end{equation}
so $\r(v_y)=0$ if and only if $\r(v_x)=0$. This analysis shows us that the Dirac velocity can be tuned to vanish by tuning $\a$ even when $C_3$ is broken. On the other hand, for $\b\neq0$ the vanishing of the Dirac velocity is not accompanied by the exact vanishing of the band dispersion. Rather, the minimal bandwidth scales linearly with $\b$ at small $\b$. The key insight for explaining the non-vanishing of the bandwidth is to notice that the analysis presented in Appendix \ref{appendix:wronskian-summary} requires the existence of  two orthogonal zero-velocity wavefunctions $\y_{a,b}$ in \eqref{nu-er} for which $\II\rarg=0$. Here, $\II$ tends to zero, but $\y_{a,b}$ become identical to one another.

When the $C_3$ symmetry is broken, the Dirac points are no longer fixed to the $K,K'$ points. In fact, to the first order in $\b$ the displacement $\d\bm k_D$ of the Dirac cones away from $K,K'$ scales as $\d\bm k_D=\mathcal{O}(\b/v_D(\a))$ and therefore \textit{diverges} near the magic angle \cite{kwan2020twisted,antebi2021inplane}. As a result of this divergence, when $\a$ is varied around the magic angle the Dirac cones travel around the BZ, meeting to form quadratic band-touching (QBT) points (see Fig. \ref{fig:Dirac points trajectory}). While the velocity (and hence $\II$) indeed vanishes at the QBT point, as required from our analysis of $\zvc$, it only happens since $\y_a$ and $\y_b$ become identical to one another. When the Dirac cones form a QBT there is only a single zero-velocity $\s_z=1$ wavefunction at this point. In that case, $\nu\rarg$ as defined in \eqref{nu-er} is constant, does not have any poles, and therefore does not guarantee any zeros of $\y_a$. \\

\section{$C_4$ symmetry breaking in the C4FB model.}
\label{app: C_4_breaking}
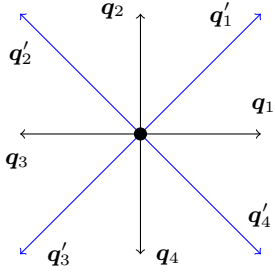
\begin{figure}
    \centering
    \begin{tikzpicture}[scale=.8]
        \foreach \i in {1,2,3,4}{
            \draw[->,black] (0,0) -- (\i*90:2cm);
            \node at ({(\i-1+.14)*90}:2.1cm) {$\bm q_\i$};
            \draw[->,blue] (0,0) -- ({(\i+1/2)*90}:2*1.414cm);
            \node at ({(\i-1/2+.12)*90}:1.414*2*.85) {$\bm q'_\i$};
        }
    \filldraw (0,0) circle (.1cm);
            
    \end{tikzpicture}
    \caption{Tunneling vectors for the simplified C4FB model \eqref{eq:simple_C4FB}. The blue and black vectors correspond to the terms in the first and second row of \eqref{eq:simple_C4FB}, respectively. }
    \label{fig:simple_C4FB}
\end{figure}
\begin{figure}
    \centering
    \begin{tikzpicture}
        \node at (0,0) {\includegraphics[scale=.8]{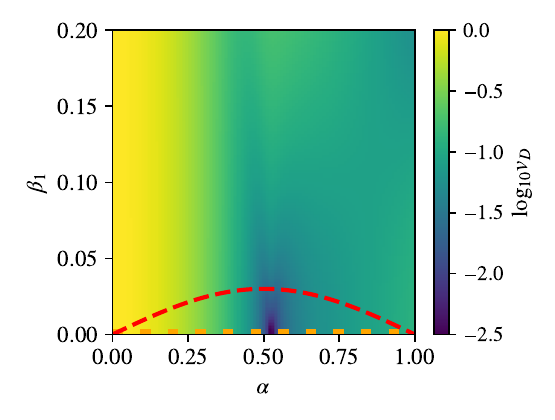}};
        \node at (-3.6,2.4) {(a)};
    \end{tikzpicture}\\
    \begin{tikzpicture}
        \node at (0,0) {\includegraphics[scale=.8]{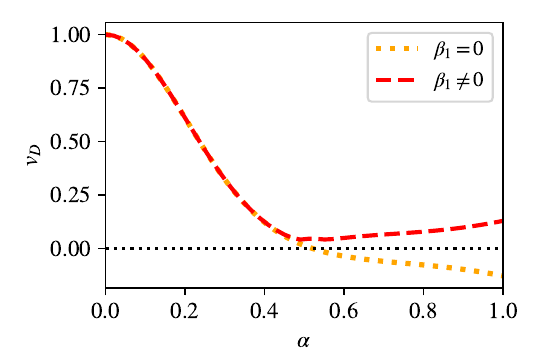}};
        \node at (-3.5,2) {(b)};
    \end{tikzpicture}\quad \quad \quad \quad
    \caption{(a) The value of the Dirac velocity at the Dirac cone for the Hamiltonian \eqref{eq:simple_C4FB} with a broken $C_4$ symmetry. The codimension of the zero Dirac velocity manifold is 2 (the dark point).
    (b) The values of the ``signed" Dirac velocity $f$ (as defined in \eqref{eq:v_decomposition_d=1}) going through two different trajectories. Since the wavefunctions are discontinuous at the magic angle the velocities acquire an opposite sign. 
    }
    \label{fig:c_4 breaking}
\end{figure}
The C4FB model \eqref{eq:H_mag_rashba_full} gives a subtle example to our analysis of $\zvc$ and in particular the theorem we proved in Appendix \ref{app:rigorous_zvc}. To simplify our analysis, we begin by constructing a continuum model with the same symmetries as the C4FB model, but which is easier to analyze. We consider
\begin{align}
    \HH &=\bm k \cdot \bm \s+\a T(\bm r) \\
    \begin{split}
    \label{eq:simple_C4FB}
    T\rarg &= \h_y\sum_n \qty(1+(-1)^n \beta_1)e^{i \bm q_n\cdot\bm r}\qty(\cos(\q_n)\s_x+\sin(\q_n)\s_y)\\
    &+\h_x\sum_n \qty(1+(-1)^n \beta_2) e^{i \bm q'_n\cdot\bm r}\qty(\cos(\q'_n)\s_x+\sin(\q'_n)\s_y)
    \end{split}
\end{align}
where  $\q_n=\qty{0,\frac{\p}{2},\p,\frac{3\p}{2}},\q'_n=\qty{\frac{\p}{4},\frac{3\p}{4},\frac{5\p}{4},\frac{7\p}{4}}$ and the inverse lattice vectors are given by $\bm q_n=(\cos\q_n,\sin\q_n),\bm q'_n=(\cos\q'_n,\sin\q'_n)$ (see Fig. \ref{fig:simple_C4FB}). The Hamiltonian has four degenerate zero-energy wave functions at ${\bm k}=0$, corresponding to two copies of the Dirac cone. 

The terms $\beta_{1,2}$ break the $C_4$ symmetry of the model to $C_2$. Notice that when either of $\beta_{1,2}$ is zero there remains a reflection symmetry. From Table \ref{tab:Double DC} we see that in the presence of any reflection symmetry we have $\zvc=1$. In the discussion of the $C_4$-symmetric model, we showed that when the model has flat bands there must be at least 8 flat bands at $E=0$ (4 per $S$ eigenvalue). A similar argument shows that in the presence of a weaker $C_2$ symmetry we must have at least 4 flat bands (2 per $S$ eigenvalue).

Interestingly, we find (Fig. \ref{fig:c_4 breaking}a) that the magic angle obtained by tuning $\a$ at $\beta_{1,2}=0$ is unstable when the $C_4$ symmetry is broken by a nonzero $\beta_{1,2}$. That is, taking $\beta_2=0,\b_1\neq 0$, for example, the codimension of the zero Dirac velocity manifold in the $\qty(\a,\b_1)$ space is not $\zvc=1$ as can naively be expected from Table \ref{tab:Double DC}. Rather, the velocity vanishes on a point in $\qty(\a,\b_1)$ space (where $\b_1=0$).

\begin{figure}
    \centering
    \includegraphics[scale=.8]{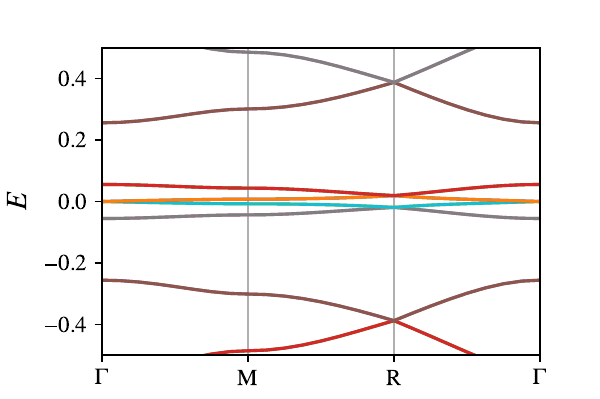}
    \caption{The band structure of the C4FB model near the parameter values at which the velocity vanishes. Each band in the picture is doubly-degenerate as a result of an antiunitary symmetry $C_2\TT'$ that squares to $-1$. We therefore find that there are 8 bands connected to $E=0$, all of which become exactly flat when the velocity vanishes.}
    \label{fig: c4 bandstructure}
\end{figure}

This apparent contradiction is resolved by noting that once either $\b_1$ or $\b_2$ are non-zero, the conditions of the theorem we proved in Appendix \ref{app:rigorous_zvc} are not satisfied, and Table \ref{tab:Double DC} cannot be used to infer the codimension. Namely, the theorem requires that there is a gap between the Dirac cone and the higher bands. In the case discussed here, for $\beta_{1,2}=0$ the $C_4$ symmetry requires that there are 8 degenerate zero-energy bands at the magic angle. Since the Dirac cone is only 4-fold degenerate, there must be 4 additional states closing the band gap (see Figure \ref{fig: c4 bandstructure}). In the presence of the $C_4$ symmetry the additional bands do not hybridize with the Dirac point wavefunctions as they have different $C_4$-eigenvalues, and we can still use the results of Appendix \ref{app:rigorous_zvc}. On the other hand, $C_4$ breaking couples the Dirac cones and higher-band wavefunctions, breaking the assumptions made in the theorem. 

A different way of understanding this phenomenon is that the definition of $f$ as in \eqref{eq:v_decomposition_d=1} is by adiabatic continuation: we always require that the wavefunctions are changed continuously as we vary the control parameters. Since the Dirac points wavefunctions are changed discontinuously at the magic angle $f$ cannot be defined to be both continuous and single-valued. As an example, in Fig. \ref{fig:c_4 breaking}b we draw $f(\alpha,\beta_1)$ going in two trajectories, one with $\beta_1=0$ and the other with nonzero $\beta_1$. The sign obtained for $f(\alpha,\beta_1)$ is opposite as a result of the discontinuity.

\section{Perturbation theory for the $C_n$-symmetric quasi-crystalline models}
\label{app: c_n perturbation}
Here we calculate in perturbation theory the formal Dirac velocity for the quasi-crystalline models \eqref{eq:c_n_matbg}. 
\subsubsection{Perturbation theory: analytical results}
Let us calculate the first orders for the perturbation series giving $v_D$. The wavefunctions are given in the form
\begin{equation}
    \y_K\rarg=\frac{1}{N}\begin{pmatrix}
    \y_{0}\rarg+\a^{2}\y_{2}\rarg+...\\
\a \y_{1}\rarg+...
\end{pmatrix}
\end{equation}
where $N$ is a normalization constant. $\y_i$ are obtained from \eqref{eq: qc wavefun}. The first terms are given by
\begin{equation}
    \begin{aligned}
    \y_0\rarg&= 1,\\
    \y_1\rarg &= -i\sum_{j=1}^n e^{i\bm q_j\cdot\bm r}, \\
    \y_2\rarg &= -\sum_{k=1}^{n-1}\sum_{j=0}^{n-1}\frac{e^{-i\qty(\bm q_{j+k}-\bm q_j)\cdot \bm r}}{1-e^{i\frac{2\p p}{n}k}},\\
    \y_3\rarg &= i\sum_{l=0}^{n-1}\sum_{k=1}^{n-1}\sum_{j=1}^{n-1}
    \frac{e^{-i\qty(\bm q_{j+k}-\bm q_j-\bm q_{j+k+l})\cdot \bm r}}{\qty(1-e^{i\frac{2\p}{n}k})\qty(1+e^{-\frac{2\p i}{n}(k+l)}-e^{\frac{2\p i}{n}l})}.
\end{aligned}
\end{equation}
The velocities are then obtained using \eqref{eq: Cn_vd} and are given in the form
\begin{equation}
    v_D= \frac{1+v_2\a^2+v_4\a^4+...}{1+N_2\a^2+N_4\a^4+...}\cdot v_0
\end{equation}
with the first coefficients given by
\begin{align}
v_2 &= -n, \\
\begin{split}
    v_4 &= n\sum_{k=1}^{n-1}\Bigg[\frac{1-2\cos\frac{2\p k}{n}+\cos\frac{4\p k}{n}}{16\sin^4\frac{\p k}{n}}\\
    &\mspace{100mu}-\frac{\cos\frac{2\p k}{n}-\cos\frac{4\p k}{n}}{2\sin^2\frac{\p k}{n}}\Bigg],
\end{split}\\
N_2 &= n, \\
N_4 &= n\sum_{k=1}^{n-1}\frac{1+2\cos\frac{2\p k}{n}-2\cos\frac{4\p k}{n}}{4\sin^2\frac{\p k}{n}}.
\end{align}
by solving for $v_D=0$ we can calculate the first magic for any $n$. Importantly, since $\zvc=1$ the coefficients of the perturbation expansion are real, which allows us to find a magic angle at finite $\a$. 
\subsubsection{Perturbation theory: numerical results}
We can numerically calculate higher-orders in the perturbation series for $n=5.$ We find
\begin{equation}
\begin{aligned}
    &v_D(n=5)= v_0\times\\
    &\frac{1-5\a^2-10\a^4-177.6\a^6-1105.5\a^8-9309.2\a^{10}+...}{1+5\a^2+20\a^4+115.2\a^6+1705.0\a^8+18841.0\a^{10}+...}
\end{aligned}
\end{equation}
which gives a magic angle at $\a_0=0.32$. 

\makeatletter\onecolumngrid@push\makeatother
\begin{table*}[!p]
	\centering
	\begin{tabular}{|cccccccccc|}
		\hline 
		$\ \Th\ $ & $\ \Pi\ $ & $\ \Si\ $ & $\ R\ $& $\Th$ rep. & $\Pi$ rep. & $\Si$ rep. & $R$ rep. & $v_{x}$ rep. & $\zvc$ \tabularnewline
		\hline 
		\hline 
		0 & 0 & 0 & $R$ & - & - & - & $\s_{y}$ & $\s_{x,z}$ & 2\tabularnewline
		&  &  & 0 & - & - & - & - & $\s_{x,y,z}$ & 3\tabularnewline
		0 & 0 & 1 & $R_{-}$ & - & - & $\s_{z}$ & $\s_{y}$ & $\s_{x}$ & 1\tabularnewline
		&  &  & 0 & - & - & $\s_{z}$ & - & $\s_{x,y}$ & 2\tabularnewline
		0 & + & 0 & $R_{+}$ & - & $\s_{x}K$ & - & $\s_{y}$ & $\s_{x}$ & 1\tabularnewline
		&  &  & 0 & - & $\s_{x}K$ & - & - & $\s_{x,y}$ & 2\tabularnewline
		- & + & 0 & $R_{-+}$ & $\s_{y}K$ & $\s_{x}K$ & $\s_{z}$ & $\s_{y}$ & $\s_{x}$ & 1\tabularnewline
		&  &  & 0 & $\s_{y}K$ & $\s_{x}K$ & $\s_{z}$ & - & $\s_{x,y}$ & 2\tabularnewline
		- & 0 & 0 & $R_{-}$ & $\s_{y}K$ & - & - & $\s_{y}$ & $\s_{x,z}$ & 2\tabularnewline
		&  &  & 0 & $\s_{y}K$ & - & - & - & $\s_{x,y,z}$ & 3\tabularnewline
		\hline 
	\end{tabular}
	\caption{\label{tab:single DC} \textbf{$\zvc$ for a single Dirac cone:} The codimension $\zvc$ of the zero-Dirac velocity manifold for a single non-degenerate Dirac cone, according to the symmetry group, with the symmetries $\Theta,\Pi,\Si,R$ defined in \eqref{eq: TPS_commutators}. In the column of each symmetry a zero denotes the absence of the symmetry, while the sign denotes the square of the symmetry. The signs of the reflection operator $R_{\zeta_{\Th},\zeta_{\Pi}}$ reflect the commutation relations of $R$ with $\Th,\Pi$. We omitted the rows that could not give rise to a single Dirac cone with the listed symmetries.\\
	For each symmetry group we specify a representation for the $\Th,\Pi,\Si,R$ operators and write matrices spanning the linear space of possible $v_x$ representations that satisfy \eqref{eq: TPS_commutators} and \eqref{eq:R_commutators}. $\zvc$ is then the dimension of this linear space.}
\end{table*}

\begin{table*}[!p]
	\centering
		\begin{tabular}{|cccccccccc|}
\multicolumn{10}{c}{}\tabularnewline
\hline 
$\ \Theta\ $ & $\ \Pi\ $ & $\ \Sigma\ $ & $\ R\ $ & $\quad\r(\Theta)\quad$ & $\quad\r\left(\Pi\right)\quad$ & $\quad\r(\Sigma)\quad$ & $\quad\r(R)\quad$ & $v_{x}$ & $\ \zvc\ $\tabularnewline
\hline 
\hline 
0 & 0 & 0 & $R$ & - & - & - & $\s_{y}$ & $\s_{x,z},\s_{x,z}\h_{x,y,z}$ & 8\tabularnewline
 &  &  & 0 & - & - & - & - & $\s_{x,y,z},\h_{x,y,z},\s_{x,y,z}\h_{x,y,z}$ & 15\tabularnewline
0 & 0 & 1 & $R_{-}$ & - & - & $\s_{z}$ & $\s_{y}$ & $\s_{x},\h_{x,y,z}\s_{x}$ & 4\tabularnewline
 &  &  & $R_{+}$ & - & - & $\h_{z}$ & $\s_{y}$ & $\s_{x,z}\h_{x,y}$ & 4\tabularnewline
 &  &  & 0 & - & - & $\s_{z}$ & - & $\s_{x,y},\s_{x,y}\h_{x,y,z}$ & 8\tabularnewline
$+$ & 0 & 0 & $R_{+}$ & $\h_{y}\s_{y}K$ & - & - & $\h_{y}\s_{y}$ & $\s_{x,z},\h_{x,z}$ & 4\tabularnewline
 &  &  & $R_{-}$ & $\h_{y}\s_{y}K$ & - & - & $\s_{y}$ & $\s_{x,z}$ & 2\tabularnewline
 &  &  & 0 & $\h_{y}\s_{y}K$ & - & - & - & $\s_{x,y,z},\h_{x,y,z}$ & 6\tabularnewline
$+$ & $+$ & 1 & $R_{++}$ & $\h_{y}\s_{y}K$ & $\s_{x}K$ & $\h_{y}\s_{z}$ & $\h_{x}\s_{y}$ & $\s_{x},\h_{y}$ & 2\tabularnewline
 &  &  & $R_{--}$ & - & - & - & - & - & -\tabularnewline
 &  &  & $R_{+-}$ & $\h_{y}\s_{y}K$ & $\s_{x}K$ & $\h_{y}\s_{z}$ & $\h_{y}\s_{y}$ & $\s_{x},\h_{x,z}$ & 3\tabularnewline
 &  &  & $R_{-+}$ & $\h_{y}\s_{y}K$ & $\s_{x}K$ & $\h_{y}\s_{z}$ & $\s_{y}$ & $\s_{x}$ & 1\tabularnewline
 &  &  & 0 & $\h_{y}\s_{y}K$ & $\s_{x}K$ & - & - & $\s_{x,y},\h_{x,z}$ & 4\tabularnewline
0 & $+$ & 0 & $R_{+}$ & - & $\s_{x}K$ & - & $\s_{y}$ & $\s_{x},\s_{z}\h_{y},\s_{x}\h_{x,z}$ & 4\tabularnewline
 &  &  & $R_{-}$ & - & $\s_{x}K$ & - & $\h_{y}$ $\s_{y}$ & $\s_{x},\s_{y}\h_{x,z},\h_{x,z},\h_{y}\s_{z}$ & 7\tabularnewline
 &  &  & 0 & - & $\s_{x}K$ & - & - & $\s_{x,y},\s_{z}\h_{y},\h_{z,x},\s_{x,y}\h_{z,x}$ & 9\tabularnewline
$-$ & $+$ & 1 & $R_{++}$ & $\h_{z}\s_{y}K$ & $\s_{x}K$ & $\s_{z}\h_{z}$ & $\h_{x}\s_{y}$ & $\s_{x},\s_{y}\h_{z}$ & 2\tabularnewline
 &  &  & $R_{--}$ & $\h_{z}\s_{y}K$ & $\s_{x}K$ & $\s_{z}\h_{z}$ & $\h_{y}\s_{y}$ & $\s_{x},\h_{x},\h_{z}\s_{y},\h_{y}\s_{z}$ & 4\tabularnewline
 &  &  & $R_{+-}$ & $\s_{y}K$ & $\s_{x}K$ & $\s_{z}$ & $\h_{y}\s_{y}$ & $\s_{x},\s_{y}\h_{x,z}$ & 3\tabularnewline
 &  &  & $R_{-+}$ & $\s_{y}K$ & $\s_{x}K$ & $\s_{z}$ & $\s_{y}$ & $\s_{x},\s_{x}\h_{x,z}$ & 3\tabularnewline
 &  &  & 0 & $\s_{y}K$ & $\s_{x}K$ & $\s_{z}$ & - & $\s_{x,y},\s_{x,y}\h_{x,z}$ & 6\tabularnewline
$-$ & 0 & 0 & $R_{+}$ & $\s_{y}K$ & - & - & $\h_{y}\s_{y}$ & $\s_{x,z}$,$\s_{y}\h_{x,z}$, $\h_{y}$ & 5\tabularnewline
 &  &  & $R_{-}$ & $\s_{y}K$ & - & - & $\s_{y}$ & $\s_{x,z},\s_{x,z}\h_{x,z}$ & 6\tabularnewline
 &  &  & $0$ & $\s_{y}K$ & - & - & - & $\s_{x,y,z},\s_{x,y,z}\h_{x,z},\h_{y}$ & 10\tabularnewline
$-$ & $-$ & 1 & $R_{++}$ & - & - & - & - & - & -\tabularnewline
 &  &  & $R_{--}$ & $\h_{x}\s_{y}K$ & $\h_{y}\s_{x}K$ & $\s_{z}\h_{z}$ & $\h_{y}\s_{y}$ & $\s_{x},\s_{z}\h_{y}$ & 2\tabularnewline
 &  &  & $R_{+-}$ & $\s_{y}K$ & $\h_{y}\s_{x}K$ & $\s_{z}\h_{y}$ & $\h_{y}\s_{y}$ & $\s_{x}$ & 1\tabularnewline
 &  &  & $R_{-+}$ & $\s_{y}K$ & $\h_{y}\s_{x}K$ & $\s_{z}\h_{y}$ & $\s_{y}$ & $\s_{x},\s_{z}\h_{x,z}$ & 3\tabularnewline
 &  &  & 0 & $\h_{x}\s_{y}K$ & $\h_{y}\s_{x}K$ & $\s_{z}\h_{z}$ & - & $\s_{x,y},\s_{z}\h_{x,y}$ & 4\tabularnewline
0 & $-$ & 0 & $R_{+}$ & - & $\h_{y}\s_{x}K$ & - & $\s_{y}$ & $\s_{x},\s_{z}\h_{x,y,z}$ & 4\tabularnewline
 &  &  & $R_{-}$ & - & $\h_{y}\s_{x}K$ & - & $\h_{y}\s_{y}$ & $\s_{x},\s_{z}\h_{y}$ & 2\tabularnewline
 &  &  & 0 & - & $\h_{y}\s_{x}K$ & - & - & $\s_{x,y},\s_{z}\h_{x,y,z}$ & 5\tabularnewline
$+$ & $-$ & 1 & $R_{++}$ & - & - & - & - & - & -\tabularnewline
 &  &  & $R_{--}$ & - & - & - & - & - & -\tabularnewline
 &  &  & $R_{+-}$ & $\h_{y}\s_{y}K$ & $\h_{y}\s_{x}K$ & $\s_{z}$ & $\h_{y}\s_{y}$ & $\s_{x}$ & 1\tabularnewline
 &  &  & $R_{-+}$ & $\h_{y}\s_{y}K$ & $\h_{y}\s_{x}K$ & $\s_{z}$ & $\s_{y}$ & $\s_{x}$ & 1\tabularnewline
 &  &  & 0 & $\h_{y}\s_{y}K$ & $\h_{y}\s_{x}K$ & $\s_{z}$ & - & $\s_{x,y}$ & 2\tabularnewline
\hline 
\end{tabular}
		\par
	\caption{\label{tab:Double DC} \textbf{$\zvc$ for two degenerate Dirac cones:} Same as Table \ref{tab:single DC} but for doubly-degenerate Dirac cones. The dashed rows signify symmetry groups which cannot support the algebra of a Dirac cone.}
\end{table*}
\clearpage
\makeatletter\onecolumngrid@pop\makeatother

\bibliography{bibliography.bib}
\end{document}